\documentclass{ws-ijmpa}
\usepackage{graphicx,graphics,epsfig,pifont}
\usepackage{latexsym}
\usepackage{txfonts}
\usepackage{hyperref}
\usepackage[english]{babel}

\RequirePackage[T1]{fontenc}
\RequirePackage[numbers]{natbib}
\usepackage{mathptmx}      
\usepackage{amssymb}      
\usepackage{amsmath}      

\newcommand{\wtb}{$Wtb~$}
\newcommand{\dzero}{$D0~$}
\newcommand{\cms}{$CMS~$}
\begin{document}

\title{ Modeling of anomalous Wtb interactions using subsidiary fields}
\author{E. Boos\footnote{boos@theory.sinp.msu.ru},
V. Bunichev\footnote{bunichev@theory.sinp.msu.ru},
L. Dudko\footnote{lev.dudko@cern.ch},
M. Perfilov\footnote{perfilov@cern.ch}}

\address{ Skobeltsyn Institute of Nuclear Physics, Lomonosov Moscow State University,
119992 Moscow, Russia.}
\maketitle

\begin{abstract}
A method to simulate anomalous fermion-boson interactions in Wtb vertex is presented with a minimal set of simulated samples of single top quark events at the LHC energies. 
In proposed method,  additional subsidiary vector fields corresponding to the Standard Model gauge fields are implemented for each type of the anomalous vertex structure.
The method allows to simulate a manifestation of anomalous gauge couplings in two approaches used in experimental analyses either keeping only the linear order 
contributions in the anomalous couplings or keeping all contributions in numerators and denominators as appeared in matrix elements. 
  For the processes with several anomalous couplings contributing simultaneously to the production and to the decay as well as to various interference terms the method allows to model correctly the dependence of kinematic distributions on anomalous couplings. The method shows how to generate a minimum set of event samples needed for a concrete analysis.
All the single top quark production mechanisms, $t$-, $s$- and associative tW-channels, are considered. The correctness of the proposed method is demonstrated.
\end{abstract}


\section{Introduction}
\label{Introduction}
  The most exciting topics of modern collider experiments are searches for physics beyond the Standard Model (BSM). ``New Physics'' may manifest itself as 
  a direct production of various new particles predicted by BSM scenarios or in a form of deviations from SM  properties  such as total, fiducial, differential cross sections and/or kinematic distributions. Such deviations could be a result of  anomalous interactions of the Standard Model (SM) particles. 
  
A well-motivated common way to introduce and model possible BSM effects below the production threshold is the 
effective Lagrangian approach in which contributions of new physics are encoded in a set of higher dimensional 
gauge invariant with respect to the SM gauge group operators  divided by an effective scale of possible new physics $1/{\Lambda}$ in corresponding power~\cite{Peccei:1989kr,Buchmuller:1985jz}.
Derivation of Feynman rules from the effective operators is straightforward. New vertices can be implemented in a computer program and necessary computations and event simulations can be performed.
However, in practice, there are a number  of mainly technical but time consuming difficulties.
For the unstable particles, such as EW gauge W and Z bosons, the Higgs boson, the top quark etc., the anomalous couplings may contribute to both production and decay, as well as to the decay width of the unstable  particle. The corresponding matrix element is then a ratio of polynomials  of anomalous parameters affecting not only the 
production rate and decay branching fraction but also spin 
correlations and corresponding distributions. Even when the leading, linear in
$1/{\Lambda}^2$, effect is analyzed, motivated by the consistency of the effective operator expansion~\cite{Zhang:2010dr}, it is quite involved to combine all the necessary terms coming from different places of the matrix element.
Several anomalous operators usually contribute  to the same vertex. Different vertexes in Feynman diagram can contain anomalous contribution, for example the production and decay vertexes of top quark. The kinematic distributions of the final particles are affected by the values of the anomalous couplings.
In order to get useful information about all  anomalous parameters one needs to simulate a large number of event samples required for experimental searches. 
This is specially problematic because of the computer time and memory consuming procedure of event propagation through detector simulation programs.
In this sense it is very reasonable to generate as small as possible number of independent event samples.
A simple practical method to simulate anomalous contributions with a minimal set of  event samples is described in this paper. The method relies on the implementation of additional subsidiary vector fields and is applicable directly for the computer programs. The simulation is performed using the CompHEP~\cite{Boos:2009un,Boos:2004kh} computer package. We demonstrate how the method works in the modeling of anomalous Wtb couplings in the single top quark production processes. 
The theory of electroweak interactions predicts three different production mechanisms for single top quarks in hadron-hadron collisions, in addition to the more abundant pair production due to the strong interaction. 
They are classified using the topology of diagrams which include W-boson~\cite{Willenbrock:1986cr} as t-channel,
 s-channel and associated tW production. Since the single top production is directly proportional 
 to the magnitude of Wtb vertex, these processes are expected to be very sensitive for the 
investigation of the Wtb vertex structure with highest possible accuracy. This property was already explored in several experimental studies at the Tevatron~\cite{Abazov:2011pm,Abazov:2008sz} and the LHC~\cite{Aad:2015yem,Khachatryan:2014vma,CMS:2014ffa}, and   recent theoretical publications~\cite{AguilarSaavedra:2011ct,Fabbrichesi:2014wva,Bernardo:2014vha,Aguilar-Saavedra:2015yza,Hioki:2015env}. In this paper we consider 
  all three production mechanisms of single top quark, model an influence of all possible  Wtb anomalous couplings with  the help of proposed method
  and validate the correctness of results by comparing with straightforward computations when the anomalous couplings are simply added to the vertex
  in the Feynman rules.
 Different approach to simulate only linear terms in $1/{\Lambda}^2$ is also possible with recent developments in MadGraph package~\cite{Alwall:2014hca}. 

\section{Method of modeling the anomalous operators in the production and decay vertices.}
\label{Idea}
The effective anomalous operators lead to changes of the SM vertices which in our case  can be written in the form of additional terms in
the SM gauge-fermion vertex:
\begin{equation}
\Gamma_{\mu} = \Gamma_{\mu}^{SM} + \Gamma_{\mu}^{NP_1} + \Gamma_{\mu}^{NP_2} + ...,
 \label{vertex}
\end{equation}
where the first term is the SM vertex of the SM gauge boson ($V_{\mu}$) and the 
SM fermions ($f$), and other terms are the anomalous ``New Physics'' parts of the vertex with 
all the allowed Lorentz structures such as $\gamma_{\mu}$, $\gamma_{\mu} \gamma_5$, $\sigma_{\mu\nu} k^{\nu}$ etc.

The main idea of the method is to introduce a new subsidiary vector bosons $V_{1~subs}^{\mu}$, $V_{2~subs}^{\mu}$ etc. 
with the  mass and all  other couplings are the same as for the SM gauge boson $V^{\mu}$, but with the couplings to the fermion $f$  being $\Gamma_{\mu}^{NP_1}$, $\Gamma_{\mu}^{NP_2}$ etc.,
respectively. In the new model there are SM boson $V^{\mu}$ with the SM model couplings and 
a number of subsidiary vector bosons $V_{1~subs}^{\mu}$, $V_{2~subs}^{\mu}$ etc., 
with the $Vff$ vertex being non-standard $\Gamma_{\mu}^{NP_1}$, $\Gamma_{\mu}^{NP_2}$ etc. All other couplings are the same as in SM. All anomalous contributions are included in the total width of the fermion $f$,
while the widths of all vector bosons including the subsidiary ones are taken to be equal to the SM gauge boson width.  
The Feynman diagram involving $V^{\mu}$ in the intermediate state will be represented in new model by several Feynman diagrams involving new bosons $V_{1~subs}^{\mu}$, $V_{2~subs}^{\mu}$ etc. The sum of all these diagrams 
is exactly the same as for the case when we simply change the SM vertex $\Gamma_{\mu}^{SM}$ 
to the new vertex $\Gamma_{\mu}$~(\ref{vertex}). The processes with bosons in the final state have to be considered with subsequent decays of the bosons to final fermions.

Within this approach it is easy to switch on or off contributions of any new subsidiary boson and compute different terms separately. This method allows to compute only the linear terms on new anomalous couplings and/or compute any interference term if needed. This is of a special importance to generate 
a minimal number of event samples.
Implementation of this method is demonstrated in the next sections. Possible anomalous contributions to Wtb vertex are computed for the single top quark production processes and most relevant distributions are presented.

\section{Anomalous \wtb~couplings in the single top quark processes.}
\label{sec:anom_wtb_in_stop_processes}

In this Section, we demonstrate how the introduced method works for the practical example related to  anomalous \wtb~couplings search in the single top quark production processes. 
The anomalous terms in Wtb vertex allowed by the Lorentz invariance are parametrized 
by the following effective Lagrangian as was proposed in~\cite{Kane:1991bg}: 
\begin{equation}
\label{anom_wtb_eq_lagrangian}
 {\cal L} =  \frac{g}{\sqrt{2}} \overline{b}{\gamma}^{\mu} \big( f_{LV} P_L + f_{RV} P_R \big) t W^{-}_{\mu} 
 + \frac{g}{\sqrt{2}}  \overline{b}~\frac{{\sigma}^{\mu\nu}  }{2 M_W} \big( f_{LT} P_L + f_{RT} P_R \big)tW^{-}_{\mu\nu}  + h.c. 
\end{equation}
Here $M_W$ is the W-boson mass, $P_{L,R} = (1 \mp \gamma_5)/2 $ is 
the left(right)-handed  projection operator, 
$W^{-}_{\mu\nu} = \partial_{\mu}W^{-}_{\nu} - \partial_{\nu}W^{-}_{\mu}$, 
$ {\sigma}^{\mu\nu} = i/2 \big[ \gamma_{\mu}, \gamma_{\nu} \big] $,
and $g$ is the weak isospin gauge coupling. Parameters 
 $f_{LV(T)}$ and  $f_{RV(T)}$ are the dimensionless coefficients 
that parametrize strengths of the left-vector (tensor) and 
the right-vector (tensor) structures in the Lagrangian. 
In the SM all fermions interact through the left-handed currents  and all constants are equal 
to zero, except $f_{LV}=V_{tb}$ (CKM-matrix element).

All the terms in the Lagrangian~(\ref{anom_wtb_eq_lagrangian})
come from various effective gauge invariant dimension-6 operators 
as given in~\cite{Whisnant:1997qu,Boos:1999ca,AguilarSaavedra:2008gt}. 
Therefore, the natural size of the strength parameters is 
of the order of $(v/{\Lambda})^2$ where $v$ is the Higgs vacuum
 expectation value and $\Lambda$ is a scale of "New Physics". 

Squared matrix element for the single top quark production processes  has a quadratic dependence on 
the anomalous constants of the following form (assuming a massless $b$-quark)~\cite{Boos:2010zzb}:
\begin{equation}
\label{xsection}
|M|^{2}_{pp \rightarrow t + X}  \sim  \ \ A \cdot f_{LV}^2 + B \cdot f_{RV}^2 
+ C \cdot f_{LV}\cdot  f_{RT}  + D \cdot f_{RV}\cdot  f_{LT} 
+ E \cdot f_{LT}^2 + G \cdot f_{RT}^2,
\end{equation}
where $A, B, C, D, E, G$ are functions of particle momenta. As one can see there are no $(f_{LV}\cdot  f_{RV})$ and $(f_{LT}\cdot  f_{RT})$ cross terms in Eq.~(\ref{xsection}). 
The contributions of the left and right parts to the squared matrix element are different in general case ($A$ is not equal to $B$, and so on) because the SM vertex of W boson with the light quarks has the left (V-A) structure. 
Simple formulas for the single top production cross sections with anomalous \wtb~couplings at the partonic level for the s-, t- and associated tW-channels  are provided in the~\ref{Appendix}. The formulas demonstrate explicitly the production cross sections dependence on the anomalous couplings being in an agreement with expressions given in~\cite{Zhang:2010dr} if only linear terms in the anomalous couplings are kept.

In the same manner as it was done in the experimental searches for the anomalous \wtb~couplings~\cite{Abazov:2008sz,Abazov:2011pm,CMS:2014ffa} we considered three scenarios with always the left-vector coupling pairing with one of  the other anomalous couplings ($(f_{LV}, f_{RV})$, $(f_{LV}, f_{LT})$, $(f_{LV}, f_{RT})$ scenarios) and other two couplings are equal to zero.
 
\begin{figure*}[!!h!]
\begin{center}
\begin{minipage}[t]{.48\linewidth}
\centering
\includegraphics[width=.98\linewidth]{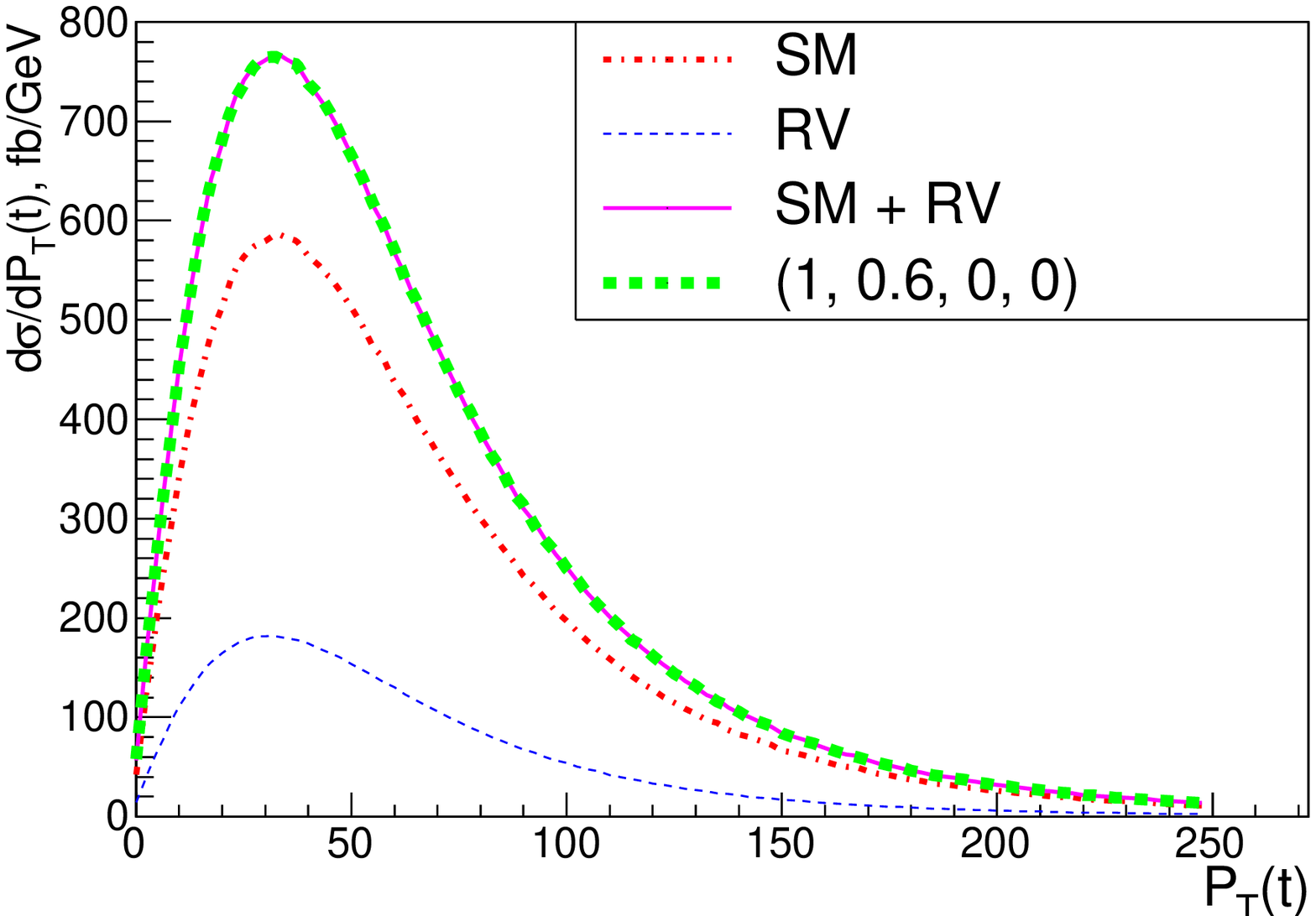}
\caption{\label{pic1}  The transverse momentum distribution of the top quark for the process $pp \rightarrow t q $ (t-channel) for the $(f_{LV}, f_{RV})$ scenario. \label{prod_vs_prod_and_decay_a}}
\end{minipage}
\hfill
\begin{minipage}[t]{.48\linewidth}
\centering
\includegraphics[width=.98\linewidth]{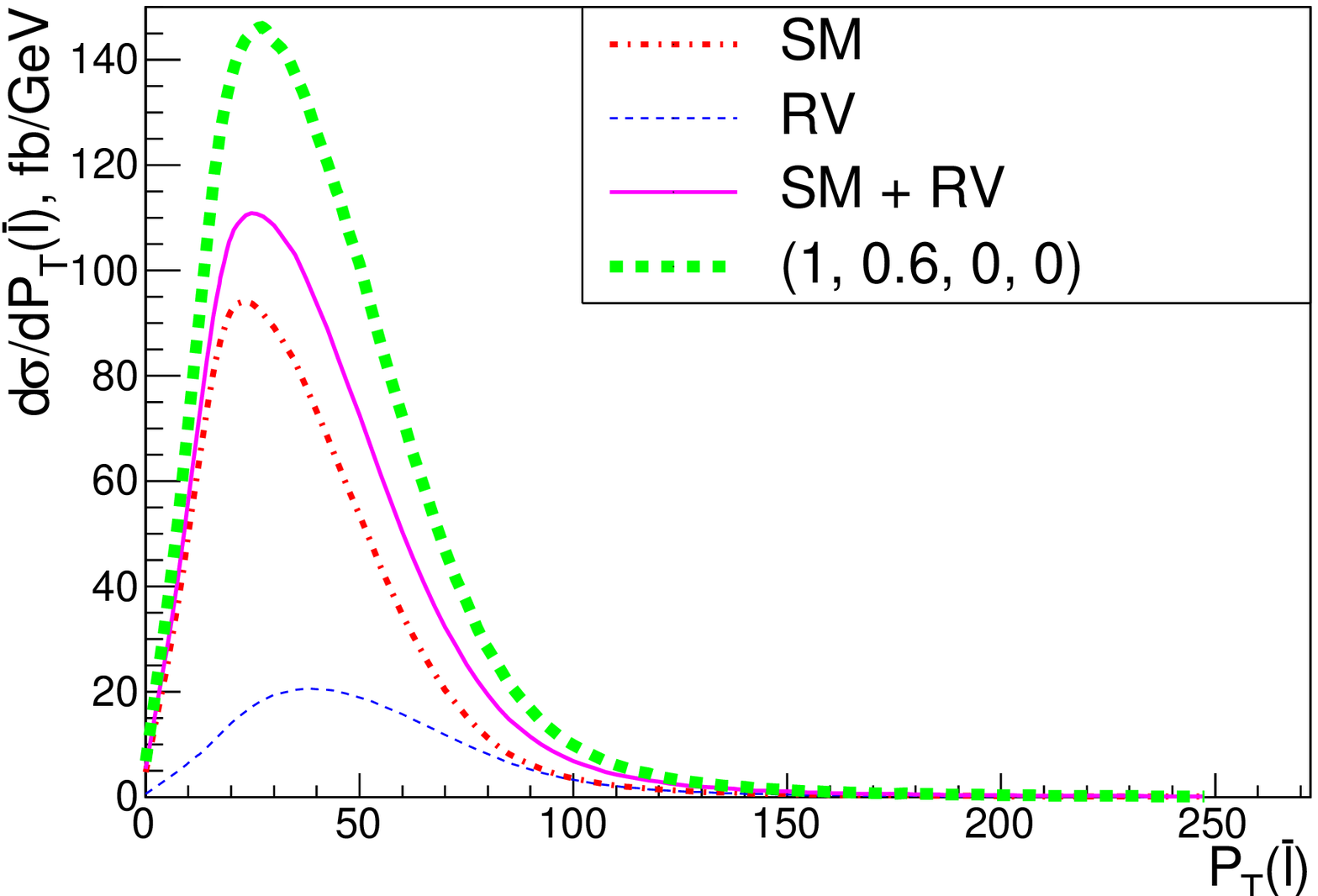}
\caption{ \label{pic2}   The transverse momentum distribution of the lepton from top quark decay for the process $pp \rightarrow t (\nu_l,\bar {l}, b)q$ for the $(f_{LV}, f_{RV})$ scenario. \label{prod_vs_prod_and_decay_b}}
\end{minipage}
\end{center}
\end{figure*}

   Let us consider briefly very simple approach when the anomalous couplings are kept only for the production part of the processes. Such an approach in the narrow width approximation is motivated by the fact that the total cross section is given by $\sigma=\sigma_{\rm production}\times BR_{\rm decay}$ and a weak dependence of the decay branching ratio $BR_{\rm decay}$ on the anomalous couplings~\cite{Boos:1999ca,Boos:1999dd}. 
The first scenario has $(f_{LV}, f_{RV})$ non-zero couplings in \wtb~vertex ($(f_{LV}, f_{RV})$ scenario). The events of the top quark production processes without top quark decay can be simulated with only two sets of events in this scenario. 
The first set of Monte-Carlo (MC) simulated events corresponds to kinematic term $A$ 
in Eq.~(\ref{xsection}) and can be obtained with the fixed values 
of the anomalous parameters: $f_{LV}=1, f_{RV}=f_{LT}=f_{RT}=0$. In this sample,
 only events with left-handed interaction in the \wtb~vertex are simulated;  the notation ``LV2''  for this set of events is used. The second set of events corresponds to the kinematic term $B$ in Eq.~(\ref{xsection}) 
and can be simulated with the 
$f_{LV}=0, f_{RV}=1$ coupling values; this set of events corresponds to the only right-handed 
interaction in the \wtb~vertex. 
The notation for such sample is ``RV2''.
As follows from Eq.~(\ref{xsection}) the kinematic distributions of a top quark production processes with all possible values of the $f_{LV}$ and $f_{RV}$ anomalous couplings can be reproduced 
by the sum of corresponding distributions following from the ``LV2'' and ``RV2'' sets of events multiplied by the squared value of the corresponding coupling. This is demonstrated in  Fig.~\ref{prod_vs_prod_and_decay_a} for the distribution of transverse momentum of the top quark for $f_{LV}=1, f_{RV}=0.6$.   The sample ``LV2'' exactly corresponds to the SM ($f_{LV}=1$) and the curve on the plots is labelled as "SM". 
The notation ``RV'' is used for the  ${f_{RV}}^2\cdot({\rm RV2}) $ curve at the plot with $f_{RV}=0.6$. One can see that the sum of the curves ``SM'' and ``RV'' is in agreement with the curve ``(1, 0.6, 0, 0)'' which shows the explicitly calculated distribution of the transverse momenta of the top quark with $f_{LV}=1, f_{RV}=0.6$ values\footnote{In order to show deviations between various curves more clearly the values for anomalous parameters are chosen significantly larger than existing experimental bounds and values followed from the SM loop contributions~\cite{GonzalezSprinberg:2011kx} to Wtb vertex as well as values~\cite{Arhrib:2016vts} predicted by the SM extensions such as 2HDM. } of the couplings in the top quark \wtb~production vertex. 
Therefore if only production is considered, then two sets of events are sufficient to perform an analysis. 
However  two sets of events are not enough if one considers the top quark production with subsequent decay. 
As one can see, for the case of $(f_{LV}, f_{RV})$ as an example, from Fig.~\ref{prod_vs_prod_and_decay_b} the distribution of the transverse momentum of the lepton from the top quark decay cannot be reproduced with a simple sum of the distributions following from ``SM'' and ``RV'' sets of events in the same way as it was done for the case of anomalous couplings in the top quark production only (Fig.~\ref{prod_vs_prod_and_decay_a}).
In the next sections it is shown that one needs to extend the minimal number of the event sets in order to correctly represent the event kinematics for the case where the anomalous couplings present in the production and decay vertices. 

\subsection{Production and decay of top quarks with anomalous couplings in $(f_{LV}, f_{RV})$ scenario}
\label{subsec:lvrv}

The matrix element squared of the single top quark production and subsequent decay  has the following structure in the narrow width approximation for the case with non-zero $f_{LV}$ and $f_{RV}$ couplings:
\begin{equation}
\label{me_sq_pr_decay}
  |M|^{2}_{\rm full}  \sim  \big ( f_{LV}^2 A_p + f_{RV}^2 B_p \big ) \frac {\big ( f_{LV}^2 A_d + f_{RV}^2 B_d \big )}{w_{\rm tot}(f_{LV},f_{RV},0,0)},
\end{equation}
where $A_p$ and $B_p$ ($A_d$ and $B_d$) are channel dependent functions of momenta in 
the production (in the decay) of the top quark, and  $w_{\rm tot}(f_{LV},f_{RV},0,0)$ is the total top quark width for the considered scenario\footnote{The total top quark width $w_{\rm tot}(f_{LV},f_{RV},f_{LT},f_{RT})$ in general case is given in~\cite{Boos:1999ca,Mohammadi Najafabadi:2006um}.}.

After a multiplication one gets the following expression:
\begin{equation}
\label{me_sq_pr_decay_1}
 |M|^{2}_{\rm full} \sim  \frac {1}{w_{\rm tot}(f_{LV},f_{RV},0,0)} \big ( f_{LV}^4 A_p A_d + f_{LV}^2 f_{RV}^2 A_p B_d  + f_{LV}^2 f_{RV}^2 A_d B_p + f_{RV}^4 B_p B_d \big ) 
\end{equation}
This expression contains couplings factorized with the kinematic functions 
from the top production and decay.
The sum of different terms in the expression reproduces the superposition of different states: 
the state with a left-handed operator in the production and in the decay 
of the top quark (first term in Eq.~(\ref{me_sq_pr_decay_1}), 
the state with right-handed operator 
in the production and the decay of top quark 
(fourth term in Eq.~(\ref{me_sq_pr_decay_1}) and the states with a left-handed 
operator in the production \wtb~vertex and a right-handed operator 
in the decay \wtb~vertex and vice versa (second and third terms 
in Eq.~(\ref{me_sq_pr_decay_1}). 

\begin{figure*}[!!h!]
\begin{center}
\includegraphics[width=.98\linewidth]{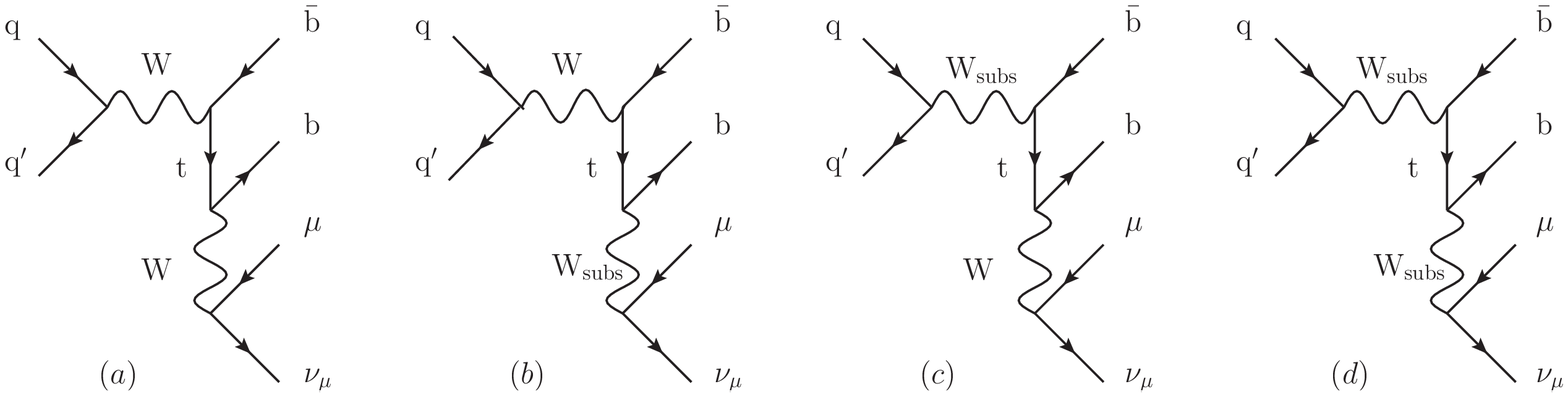}
\caption{Diagrams for s-channel single top quark production 
with the additional subsidiary vector field with the same properties as the SM W boson but
 having the right-handed interaction in the \wtb~vertex \label{fig:4diags}
}
\end{center}
\end{figure*}

In practice the introduction of an subsidiary vector charged boson ($W_{\rm subs}$) with 
the same properties as the SM W boson but with the right-handed (anomalous) 
interaction in the \wtb~vertex (as described in Sec.~\ref{Idea}) 
makes possible to simulate the second, third and fourth terms of Eq.~(\ref{me_sq_pr_decay_1}).
Feynman diagrams for all contributions including the SM diagram (~\ref{fig:4diags}(a))
and three additional diagrams with the new subsidiary field $W_{\rm subs}$ are shown in Fig.~\ref{fig:4diags}.
Diagrams (b) and (c) correspond to the second and third terms of Eq.~(\ref{me_sq_pr_decay_1}),
the diagram (d) corresponds to the fourth term in Eq.~(\ref{me_sq_pr_decay_1}) describing 
purely right-handed interaction in \wtb vertex. 

If needed for various analyses one can combine or exclude some diagrams. For example, if one is interested in study of 
the only leading order effects in the anomalous coupling, corresponding to the leading order in  $1/{\Lambda^2}$, 
one can keep only the diagram (a) squared and the interferences of the diagram (a) with the diagrams (b) and (c) and remove all the other squared diagrams.  

If all the contributions are kept, MC simulation of the events with ($f_{LV}, f_{RV}$) couplings  requires three sets of the event samples. The first set of simulated events corresponds to left-handed interactions represented by the diagram \ref{fig:4diags}(a) (notation for the sample ``LV4''). The second set corresponds to right-handed interactions represented by the diagram  \ref{fig:4diags}(d) (notation for the sample ``RV4'').
The third set of simulated events corresponds to left-handed interaction in the 
top quark production vertex and right-handed interaction in the top quark
decay vertex, and vice versa (diagrams  \ref{fig:4diags}(b),(c)) 
(notation for the sample ``LV2RV2'').

The final expression for the simulated event samples combination is:
\begin{equation}
(f_{LV},~f_{RV},~0,~0) = f_{LV}^4\cdot({\rm LV4 }) + f_{LV}^2 f_{RV}^2 \cdot ({\rm LV2RV2 }) + f_{RV}^4\cdot({\rm RV4 }),
\label{sigma_lvrv_modelling}
\end{equation}
Eq. (\ref{sigma_lvrv_modelling}) and similar equations below for
combinations of generated event samples  mean that any common kinematic
distribution with certain values of anomalous couplings is the sum of
corresponding distributions extracted from each individual event sample,
such as "LV4","LV2RV2" and "RV4" in case of
Eq.(\ref{sigma_lvrv_modelling}), multiplied by the corresponding
coefficients such as $f_{LV}^4$, $f_{LV}^2 \cdot f_{RV}^2$ and $f_{RV}^4$.

\begin{figure*}[htb]
\centering
\includegraphics[width=.45\textwidth]{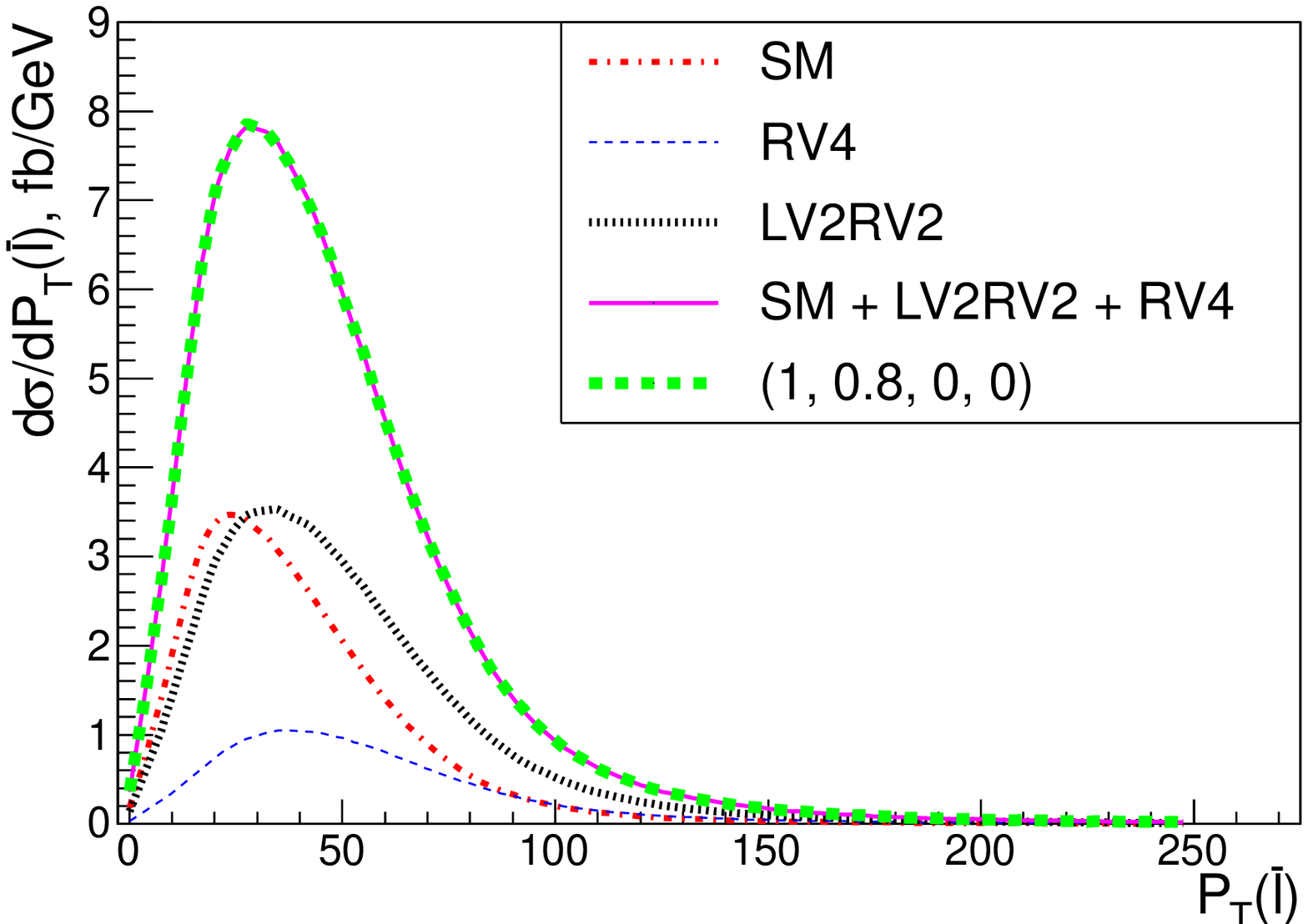}
\includegraphics[width=.45\textwidth]{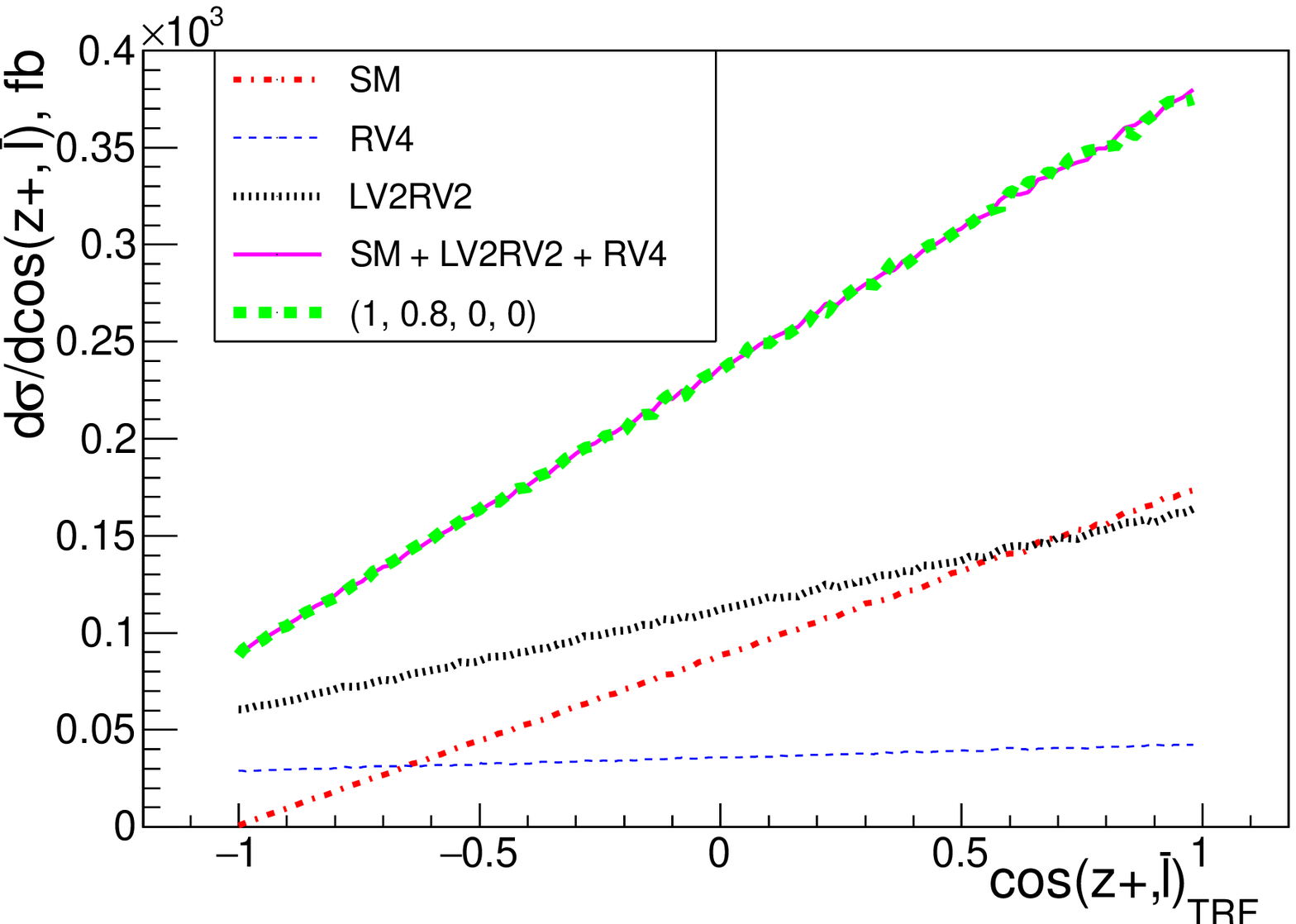}
\caption { The distributions for the transverse momentum of lepton from top quark decay in center of mass rest frame (left plot) and the cosine of angle between lepton from W from top quark decay and the down-type quark in the initial state (right plot) for the process $pp \rightarrow t (\nu_{l},\bar {l}, b) \bar b $ (s-channel single top quark production) for $(f_{LV}, f_{RV})$ scenario. \label{LVRV_s_ch_plots} }
\end{figure*}

\begin{figure*}[htb]
\centering
\includegraphics[width=.45\textwidth]{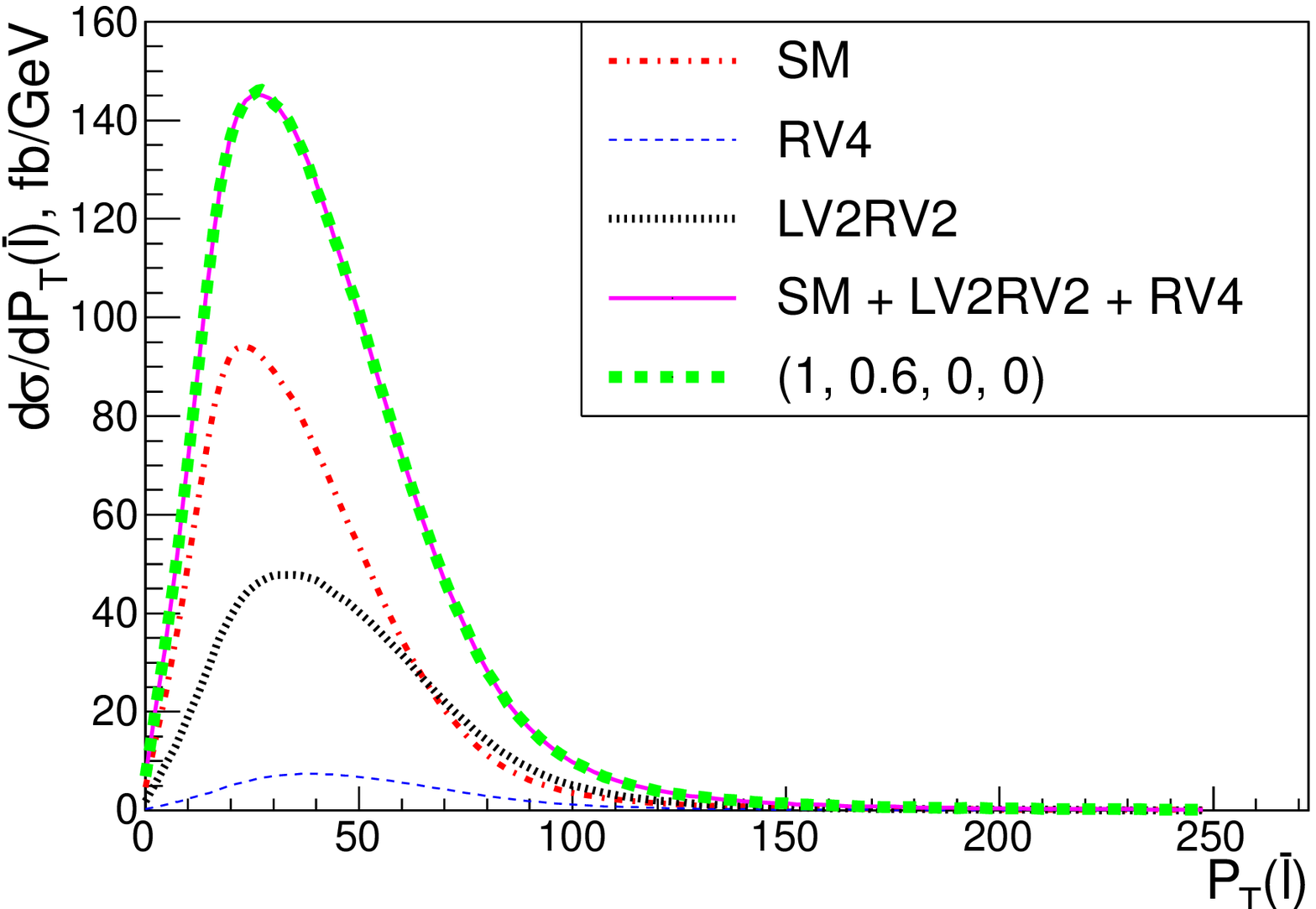}
\includegraphics[width=.45\textwidth]{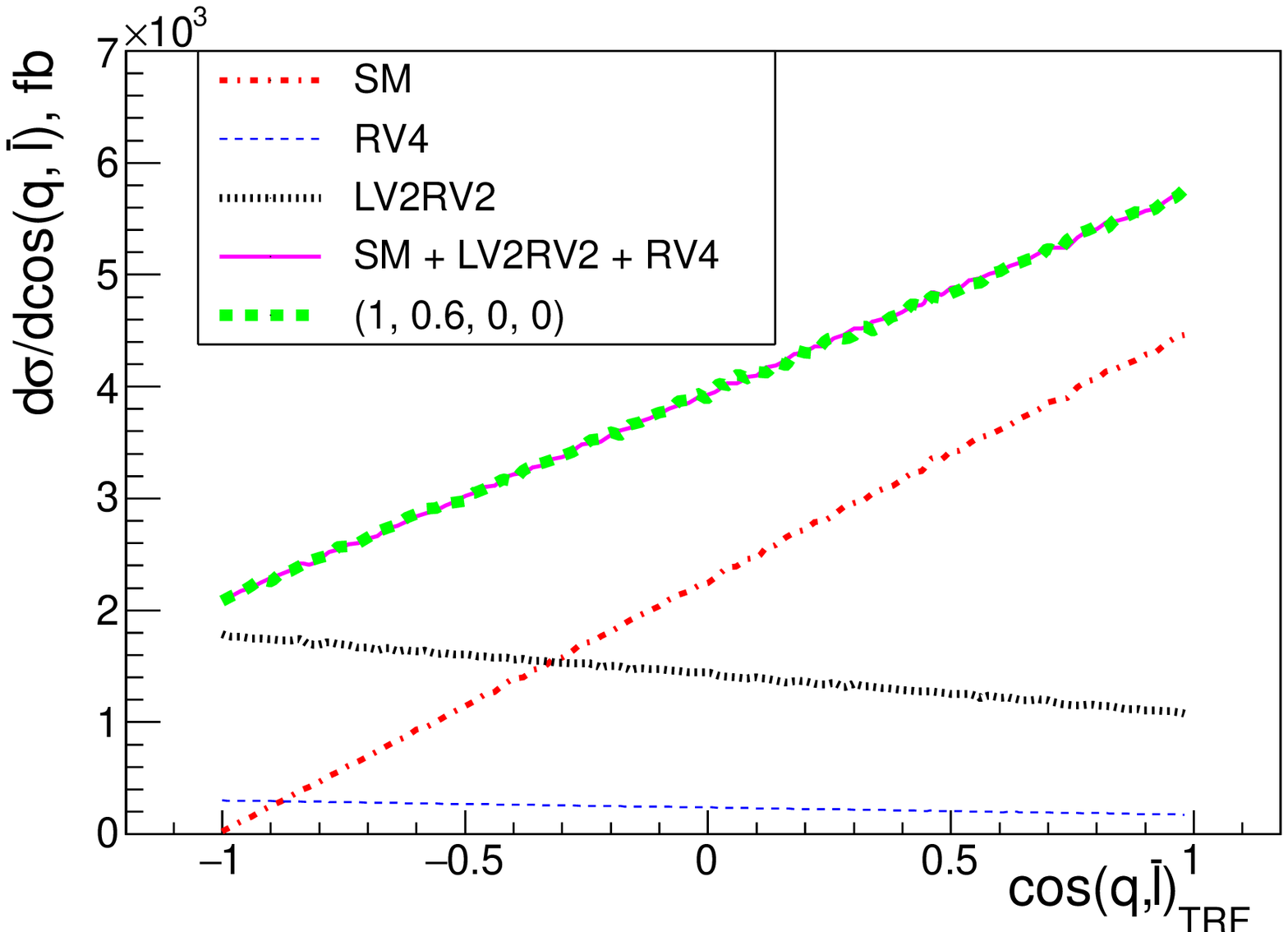}
\caption { The distributions for the transverse momentum of lepton from top quark decay in center of mass rest frame (left plot) and the cosine of angle between lepton from the top quark decay and light quark in the top quark rest frame (right plot) for the process $pp \rightarrow t (\nu_{l},\bar {l}, b) q$ (t-channel single top quark production) for $(f_{LV}, f_{RV})$ scenario. \label{LVRV_t_ch_plots} }
\end{figure*}

\begin{figure*}[htb]
\centering
\includegraphics[width=.45\textwidth]{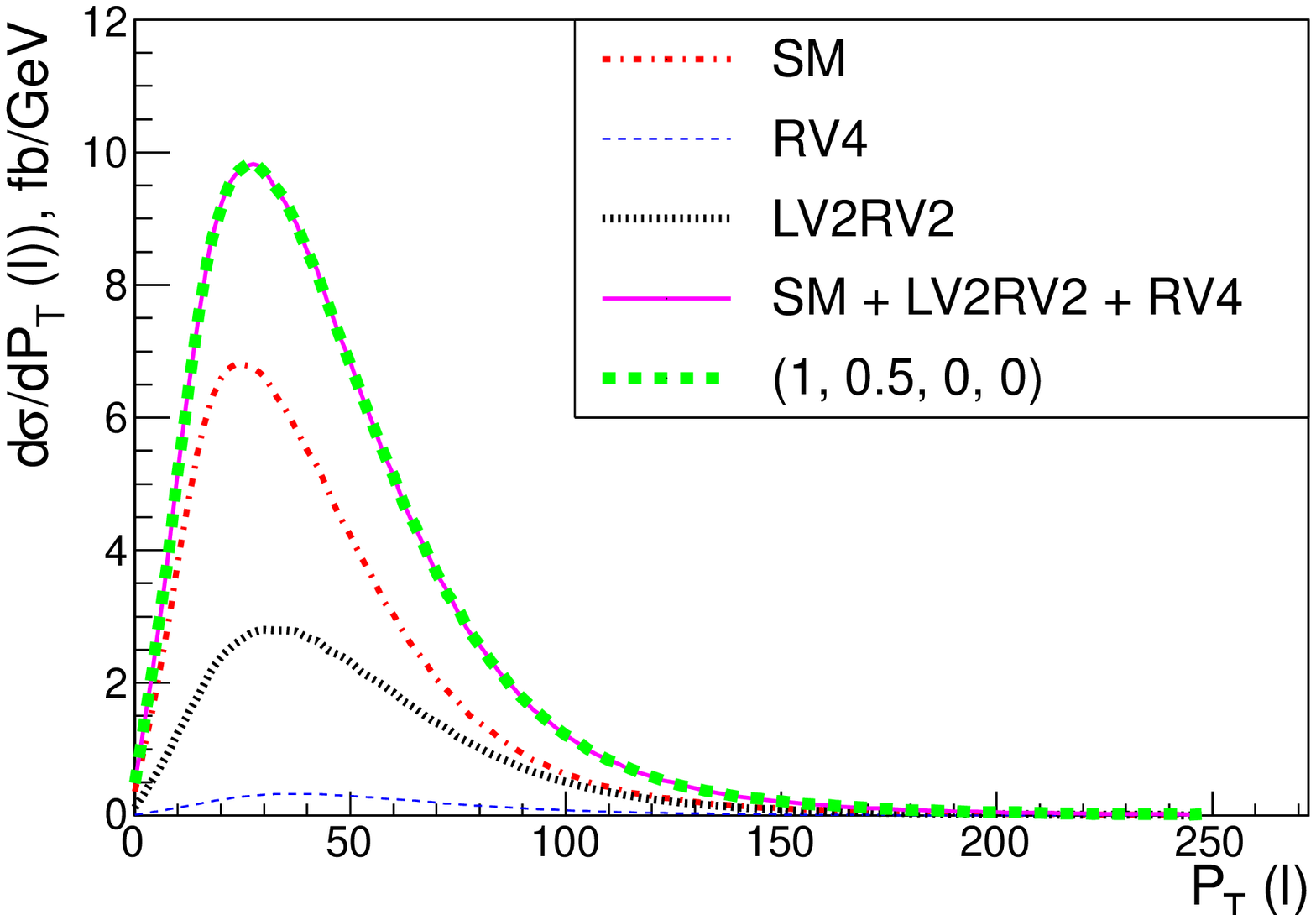}
\includegraphics[width=.45\textwidth]{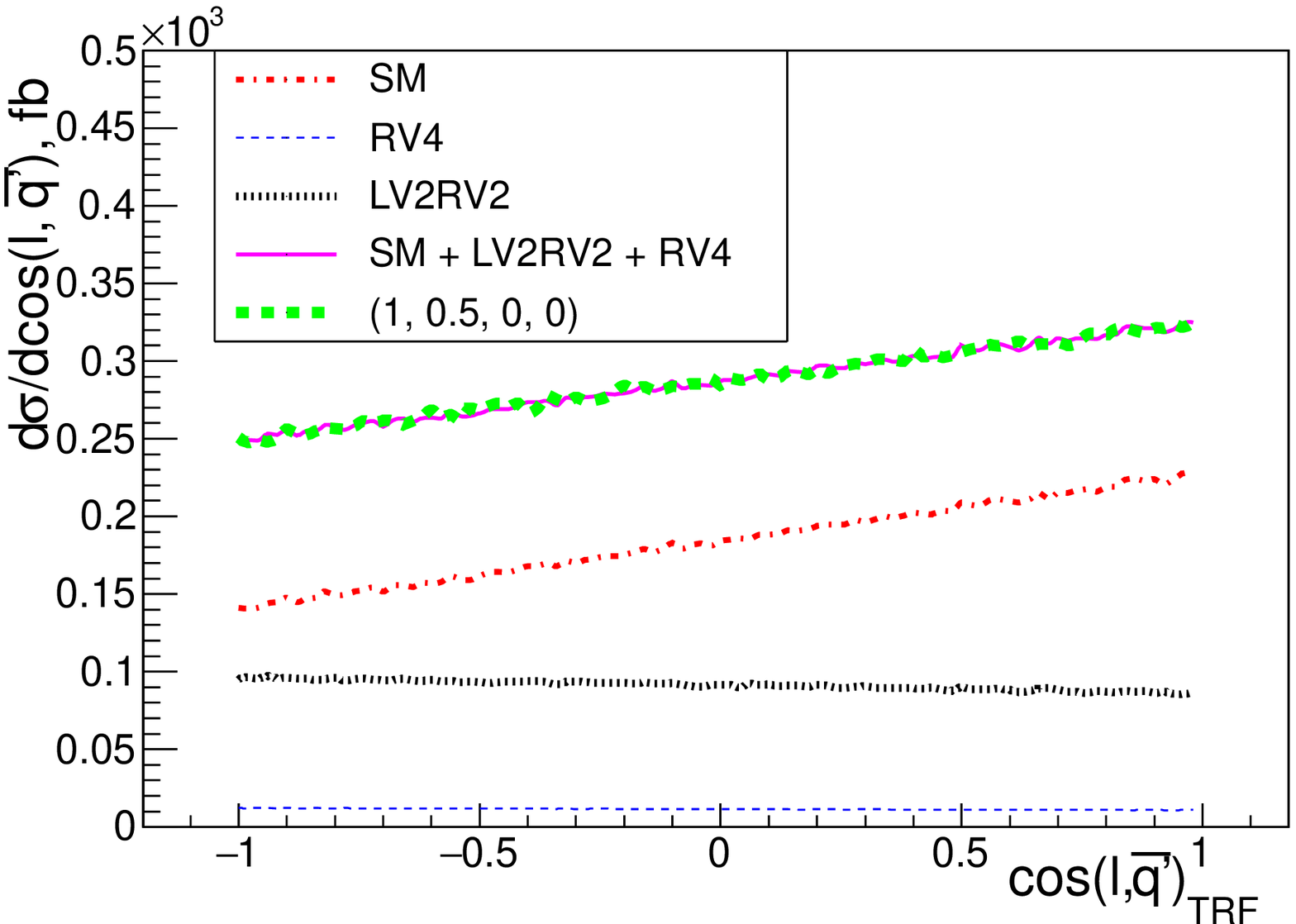}
\caption { The distributions for the transverse momentum of lepton from top quark decay in center of mass rest frame (left plot) and the cosine of angle between lepton from the top quark decay and down-type quark in the top quark rest frame (right plot) for the process $pp \rightarrow t (\bar {\nu_{l}},{l}, \bar {b}), q, \bar {q'} $ (tW-channel single top quark production) for $(f_{LV}, f_{RV})$ scenario. \label{LVRV_tW_ch_plots} }
\end{figure*}

The ability of the method to simulate different anomalous contributions for the ($f_{LV}$ , $f_{RV}$) scenario was tested for
many values of anomalous Wtb couplings. Examples are given 
in Fig.~\ref{LVRV_s_ch_plots} for the s-channel ($f_{LV} = 1.0$, $f_{RV} = 0.8$), in Fig.~\ref{LVRV_t_ch_plots} for the t-channel ($f_{LV} = 1.0$, $f_{RV} = 0.6$), and in Fig.~\ref{LVRV_tW_ch_plots} for the associated tW single top quark production ($f_{LV} = 1.0$, $f_{RV} = 0.5$).  As one can see in all cases a very good agreement between the sum 
of ``SM'', ``LV2RV2'', and ``RV4'' contributions and the straightforward 
complete matrix element computations involving anomalous parameters in the Wtb vertex 
is found (since in all examples the SM value is used the parameter $f_{LV}$
the notations "SM" in the figures is taken for the distributions obtained from the sample  "LV4").  In Fig.~\ref{LVRV_s_ch_plots}-\ref{LVRV_tW_ch_plots} such an agreement is demonstrated for the distribution of the lepton transverse momenta and the cosine of angle between lepton from top quark decay and the down-type quark in the initial state (for $s$-channel),
the light quark~\cite{Mahlon:1996pn}  (for $t$-channel) and the down-type quark~\cite{Boos:2002xw} (for associated $tW$-channel) in the top quark rest frame. The described variables are chosen  as examples of usually used characteristic variables in phenomenological and experimental analyses of the single top quark production and used throughout the entire paper. One could clearly see that presence of left- and right-handed vector operators in \wtb vertices is modeled by three sets of events which  correspond to the first (``LV4''), second (``LV2RV2'') and third (``RV4'') terms of Eq.~(\ref{sigma_lvrv_modelling}). 


\subsection{$(f_{LV}, f_{LT})$ scenario}
\label{subsec:lvlt}
The case with left vector $f_{LV}$ and left tensor $f_{LT}$ non-zero couplings in \wtb~vertex is similar to the one described 
in Sec. \ref{subsec:lvrv} because the cross term with $f_{LV}$ and $f_{LT}$ 
multiplication is absent in (\ref{xsection}). The three sets of 
events are needed for accurate simulation of the kinematics in the ($f_{LV}, f_{LT}$) scenario.
The first set of events corresponds to the diagram \ref{fig:4diags}(a) (notation for the sample ``LV4'') with left-handed vector interaction. The second set of simulated events with pure anomalous left-handed tensor
 interaction in the \wtb~vertex (with $f_{LV}=f_{RV}=f_{RT}=0, f_{LT}=1$), corresponds to the diagram from Fig.~\ref{fig:4diags}(d) (notation for the sample ``LT4''). The third set of simulated events is related to the diagrams  from Fig.~\ref{fig:4diags}(b),(c) (notation for the sample ``LV2LT2'') and corresponds to left-handed vector interaction in the top quark production vertex and left-handed tenzor coupling in the top quark decay vertex, and vice versa.
The combined expression is similar to that in Eq.~(\ref{sigma_lvrv_modelling}):
\begin{equation}
(f_{LV},~0,~f_{LT},~0) =  f_{LV}^4 \cdot ({\rm LV4}) + f_{LV}^2 f_{LT}^2 \cdot ({\rm LV2LT2}) 
 +  f_{LT}^4 \cdot  ({\rm LT4}),
\label{sigma_lvlt_modelling}
\end{equation}

\begin{figure*}[h]
\centering
\includegraphics[width=.45\textwidth]{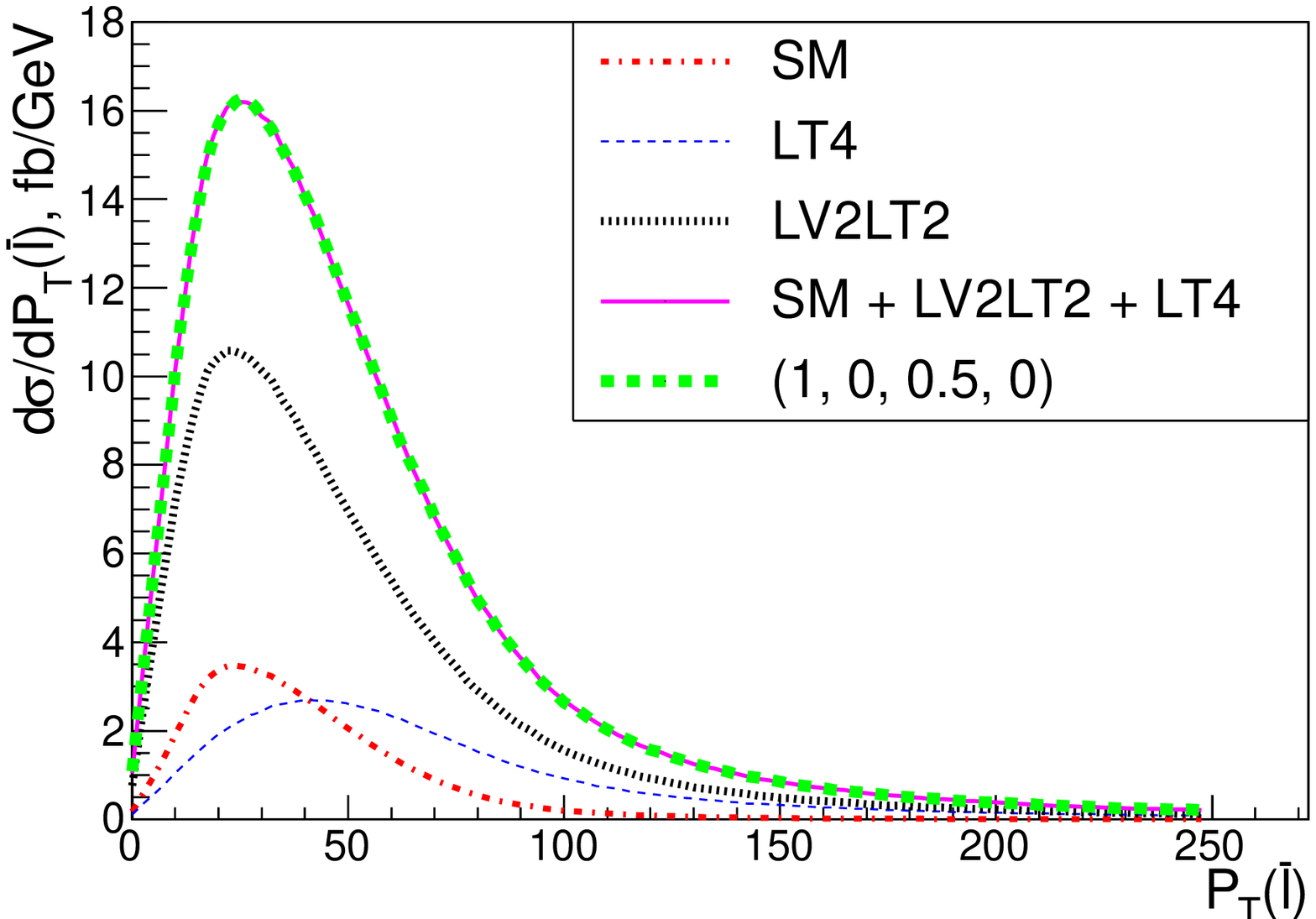}
\includegraphics[width=.45\textwidth]{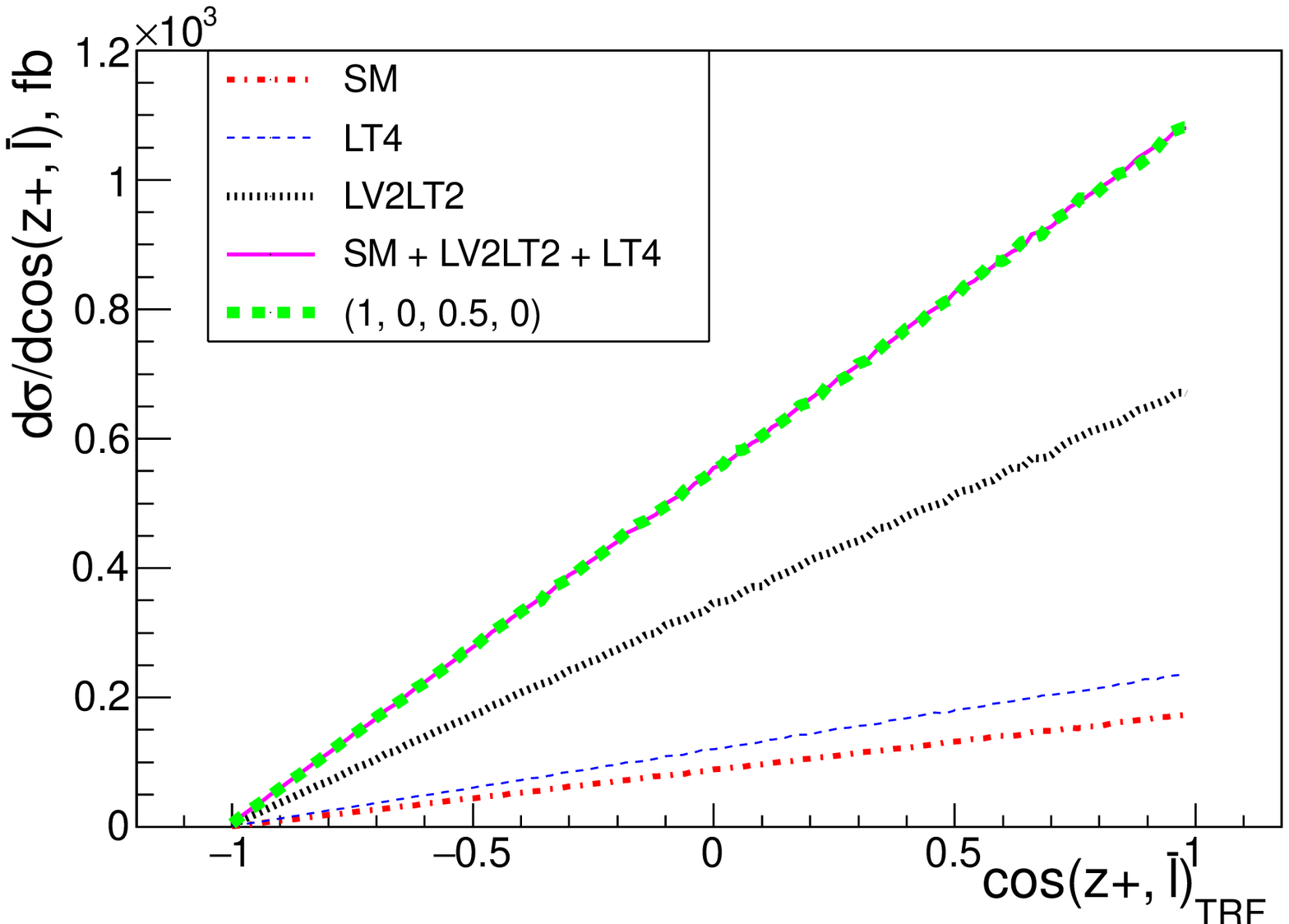}
\caption { The distributions for the transverse momentum of the lepton (left plot) and the cosine of angle between lepton from the W boson decay from top quark and the down-type quark in the initial state (right plot) for the process $pp \rightarrow t (\nu_{l},\bar {l}, b) \bar b $ (s-channel single top quark production) for $(f_{LV}, f_{LT})$ scenario. \label{LVLT_s_ch_plots} }
\end{figure*}

\begin{figure*}[h]
\centering
\includegraphics[width=.45\textwidth]{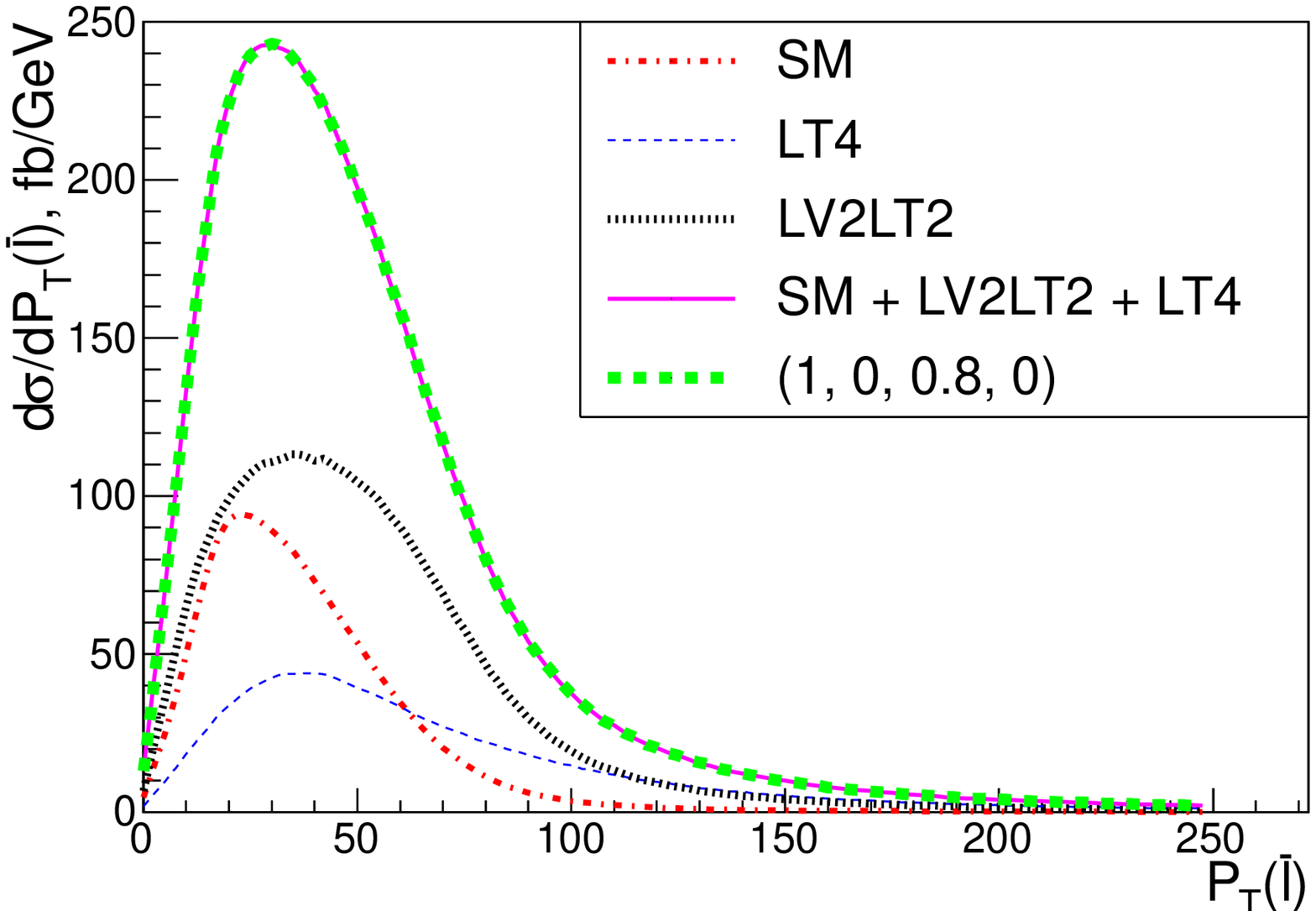}
\includegraphics[width=.45\textwidth]{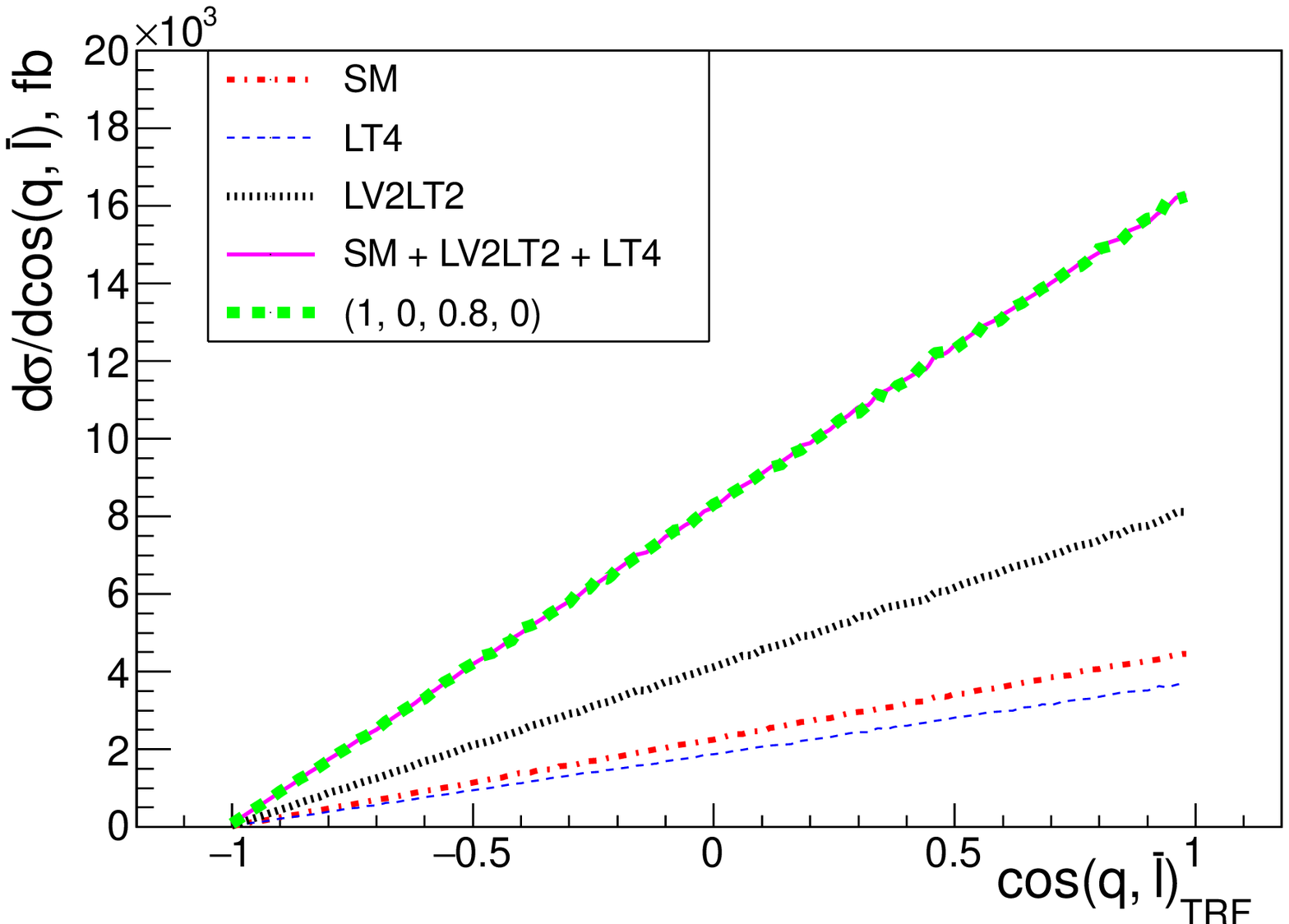}
\caption { The distributions for the transverse momentum of the lepton (left plot) and the cosine of angle between lepton from the W boson decay from top quark and light quark in the top quark rest frame (right plot) for the process $pp \rightarrow t (\nu_{l},\bar {l}, b) q$ (t-channel single top quark production) for $(f_{LV}, f_{LT})$ scenario.  \label{LVLT_t_ch_plots} }
\end{figure*}

\begin{figure*}[h]
\centering
\includegraphics[width=.45\textwidth]{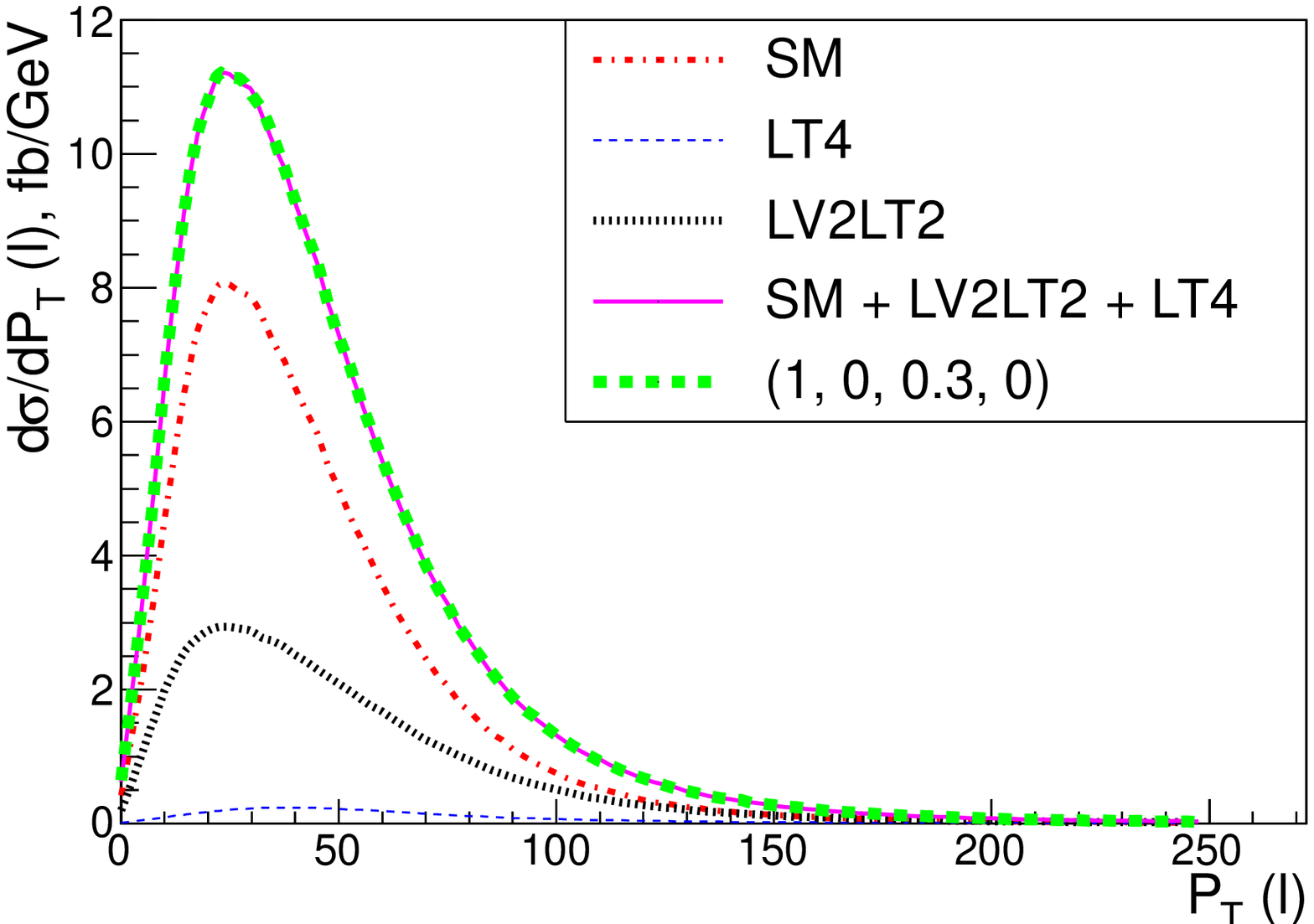}
\includegraphics[width=.45\textwidth]{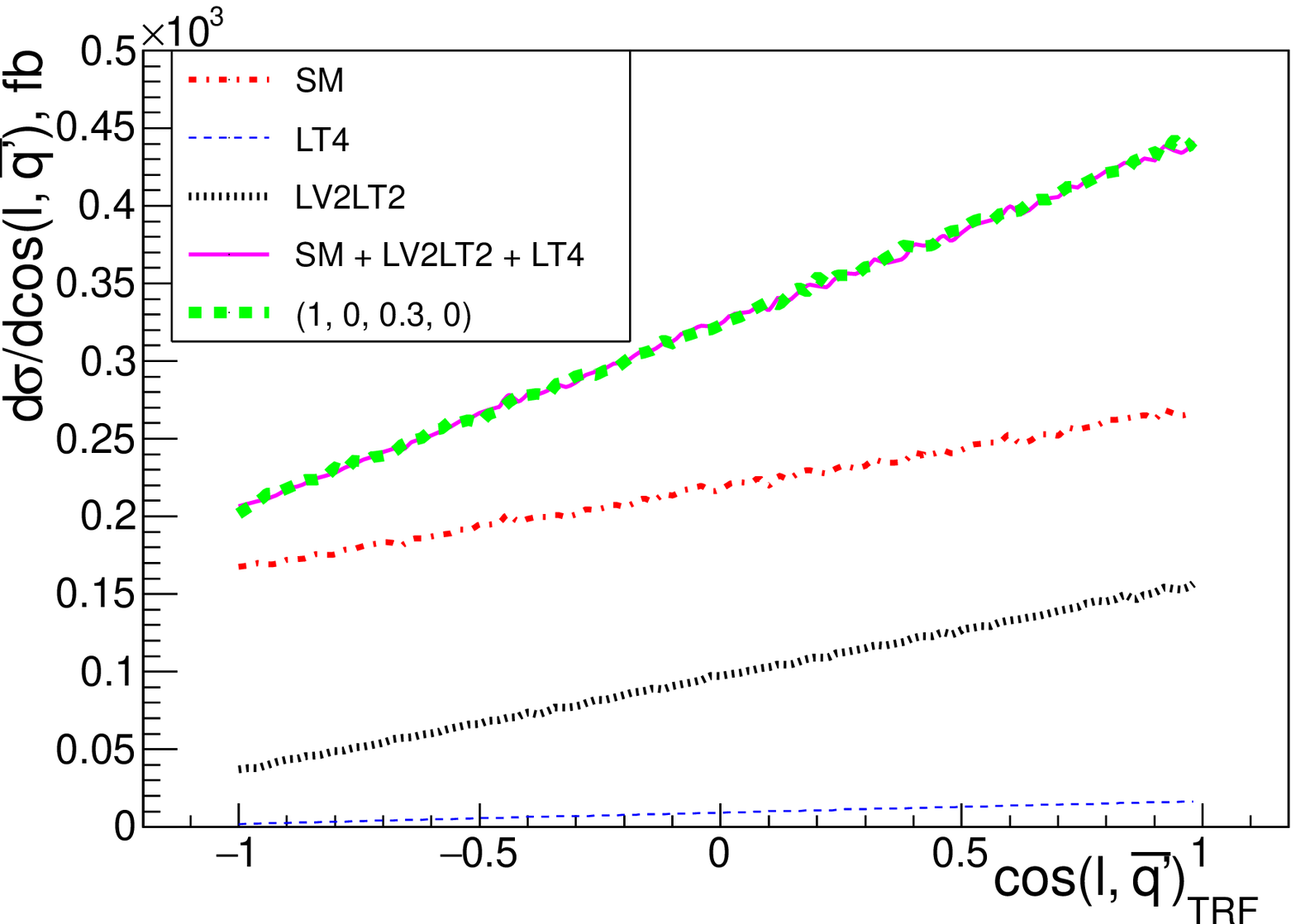}
\caption {  The distributions for the transverse momentum of lepton from top quark decay in center of mass rest frame (left plot) and the cosine of angle between lepton from the top quark decay and down-type quark in the top quark rest frame (right plot) for the process $pp \rightarrow t (\bar {\nu_{l}},{l}, \bar {b}), q, \bar {q'} $ (tW-channel single top quark production) for $(f_{LV}, f_{LT})$ scenario.   \label{LVLT_tW_ch_plots} }
\end{figure*}

In Fig.~\ref{LVLT_s_ch_plots}  one can see that sum of the curves ``SM'', ``LV2LT2'' and ``LT4'' on the plots (which are represent the first, second, and third terms in Eq.~\ref{sigma_lvlt_modelling} respectively) is in agreement with the curve ``1, 0, 0.5, 0'' for the s-channel distributions described in Sec. \ref{subsec:lvrv}. The same agreement is demonstrated in Fig.~\ref{LVLT_t_ch_plots} and  \ref{LVLT_tW_ch_plots} for two other channels of the single top quark production and  $f_{LV} = 1, f_{LT}=0.8$ and  $f_{LV} = 1, f_{LT}=0.3$ values of the couplings.

\subsection{$(f_{LV}, f_{RT})$ scenario}
\label{subsec:lvrt}
The scenario with $f_{LV}$ and $f_{RT}$ couplings  in \wtb~vertex is the more complicated than the ones described in previous sections  
\ref{subsec:lvrv} and \ref{subsec:lvlt} due to the presence 
of the ($f_{LV} \cdot f_{RT}$) cross term in Eq.~(\ref{xsection}).

In this scenario five kinematic terms with different 
powers of constants $f_{LV}$ and $f_{RT}$ arise if one considers the squared matrix element:
\begin{equation}
\label{me_sq_lvrt_pr_decay}
|M|^{2}_{\rm full} \sim \sum^{4}_{i=0} k_{i} \cdot P_{4-i} D_i 
\end{equation}
where $P,D$ are some kinematic functions of the production and decay 
of top quarks and 
$k_{i}=f_{LV}^{4-i} f_{RT}^{i} $ (upper $i$ is power, not an index).

The idea is to combine the event samples which are related to the same powers of constants in~(\ref{me_sq_lvrt_pr_decay}). At the computational level it means the selection of the squared diagrams  for the process of single top quark production and the subsequent decay with the SM W boson and $W_{\rm subs}$; the last one has the right-handed tenzor coupling with top and bottom quarks in the \wtb~vertex and the SM-like values of couplings in all other vertices. For example the term $k_{1}=(f_{LV})^{3} (f_{RT})^{1}\cdot P_3 D_1$ from Eq.~(\ref{me_sq_lvrt_pr_decay})
corresponds to the set of squared diagrams which have three W bosons and one $W_{\rm subs}$ boson.

The following minimal set of simulated event samples is needed for correct reproduction of the kinematics for the case of arbitrary values of $f_{LV}$, $f_{RT}$ couplings are in the \wtb~vertex. The ``LV4'' event sample with pure left-handed vector interaction (with $f_{LV}=1, f_{RV}=f_{LT}=f_{RT}=0$) is represented by diagram from Fig.~\ref{fig:4diags}(a).  The ``RT4'' event sample with pure right-handed tensor interaction (with $f_{LV}=f_{RV}=f_{LT}=0, f_{RT}=1$) is represented by diagram from Fig.~\ref{fig:4diags}(d). Three additional event samples ``LV3RT1'', ``LV2RT2'', ``LV1RT3'' are related to the cross terms with corresponding powers of the couplings $f_{LV}$ and $f_{RT}$ in Eq.~(\ref{me_sq_lvrt_pr_decay}).

\begin{figure*}[h]
\centering
\includegraphics[width=.45\textwidth]{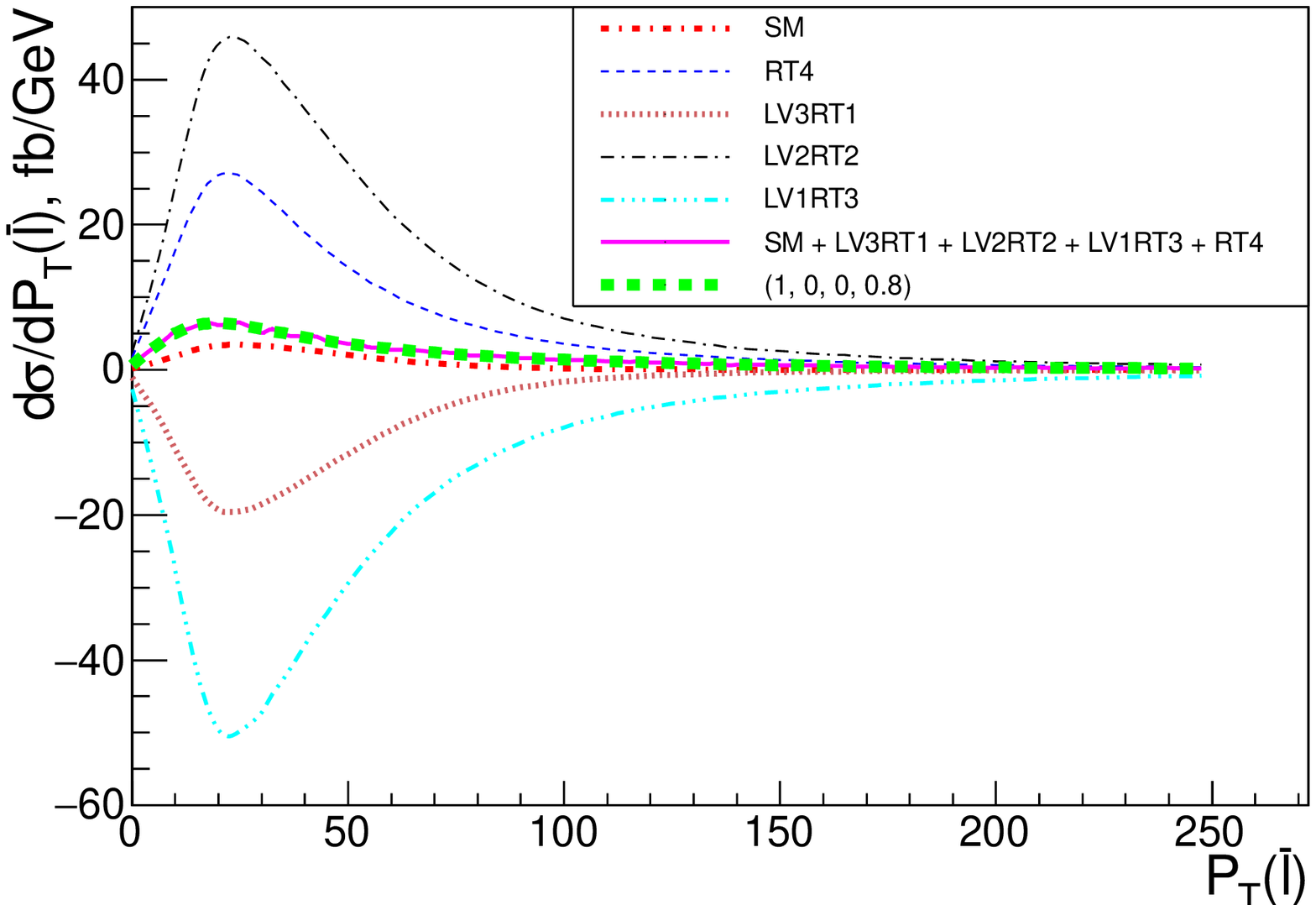}
\includegraphics[width=.45\textwidth]{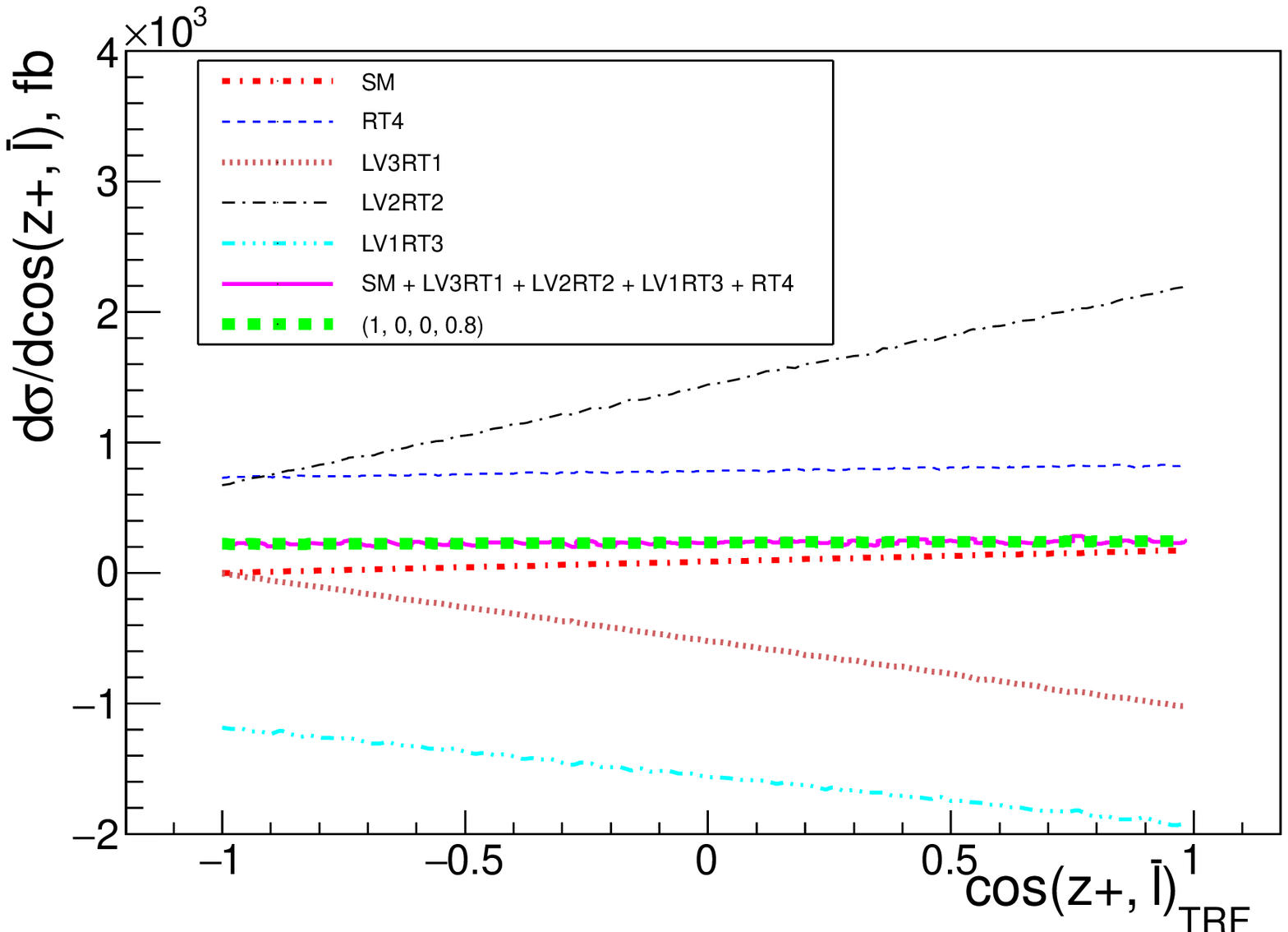}
\caption { The distributions for the transverse momentum of lepton from top quark decay in center of mass rest frame (left plot) and the cosine of angle between lepton from top quark decay and the down-type quark in the initial state (right plot) for the process $pp \rightarrow t (\nu_{l},\bar {l}, b) \bar b $ (s-channel single top quark production) for $(f_{LV}, f_{RT})$ scenario. \label{LVRT_s_ch_plots} }
\end{figure*}

\begin{figure*}[h]
\centering
\includegraphics[width=.45\textwidth]{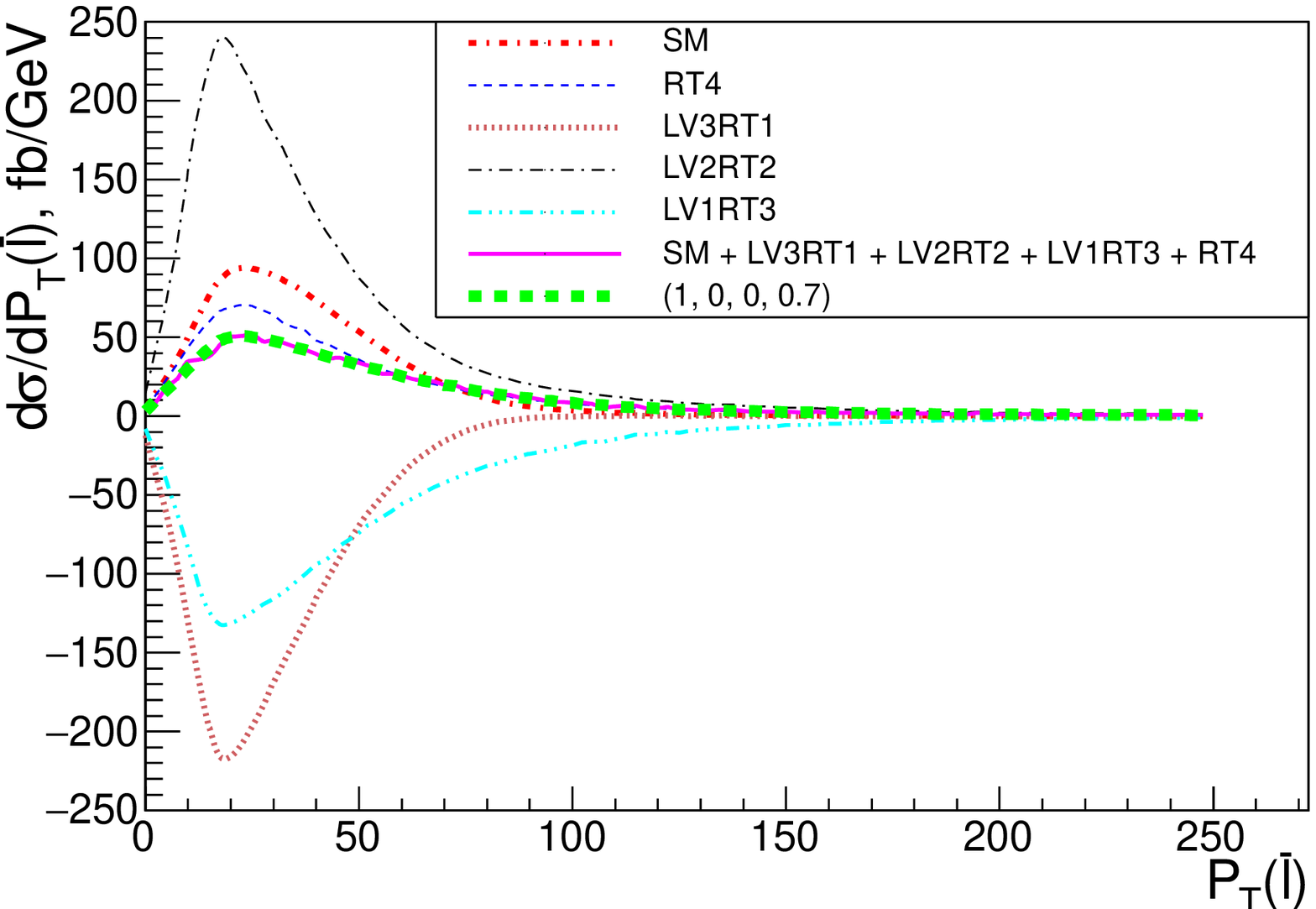}
\includegraphics[width=.45\textwidth]{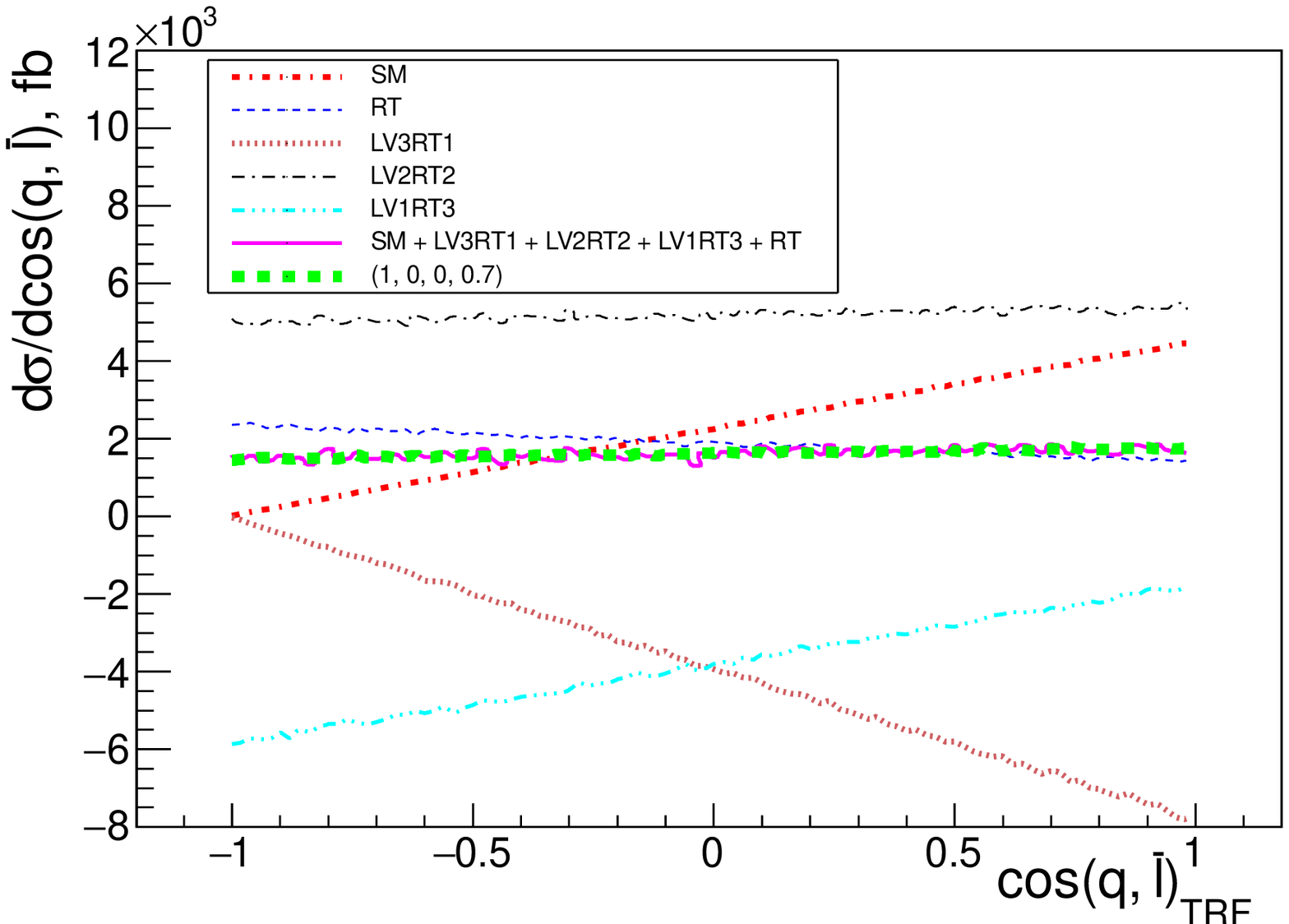}
\caption { The distributions for the transverse momentum of top quark in center of mass rest frame (left plot) and the cosine of angle between lepton from the W boson decay from top quark and light quark in the top quark rest frame (right plot) for the process $pp \rightarrow t (\nu_{l},\bar {l}, b) q$ (t-channel single top quark production) for $(f_{LV}, f_{RT})$ scenario. \label{LVRT_t_ch_plots} }
\end{figure*}

\begin{figure*}[h]
\centering
\includegraphics[width=.45\textwidth]{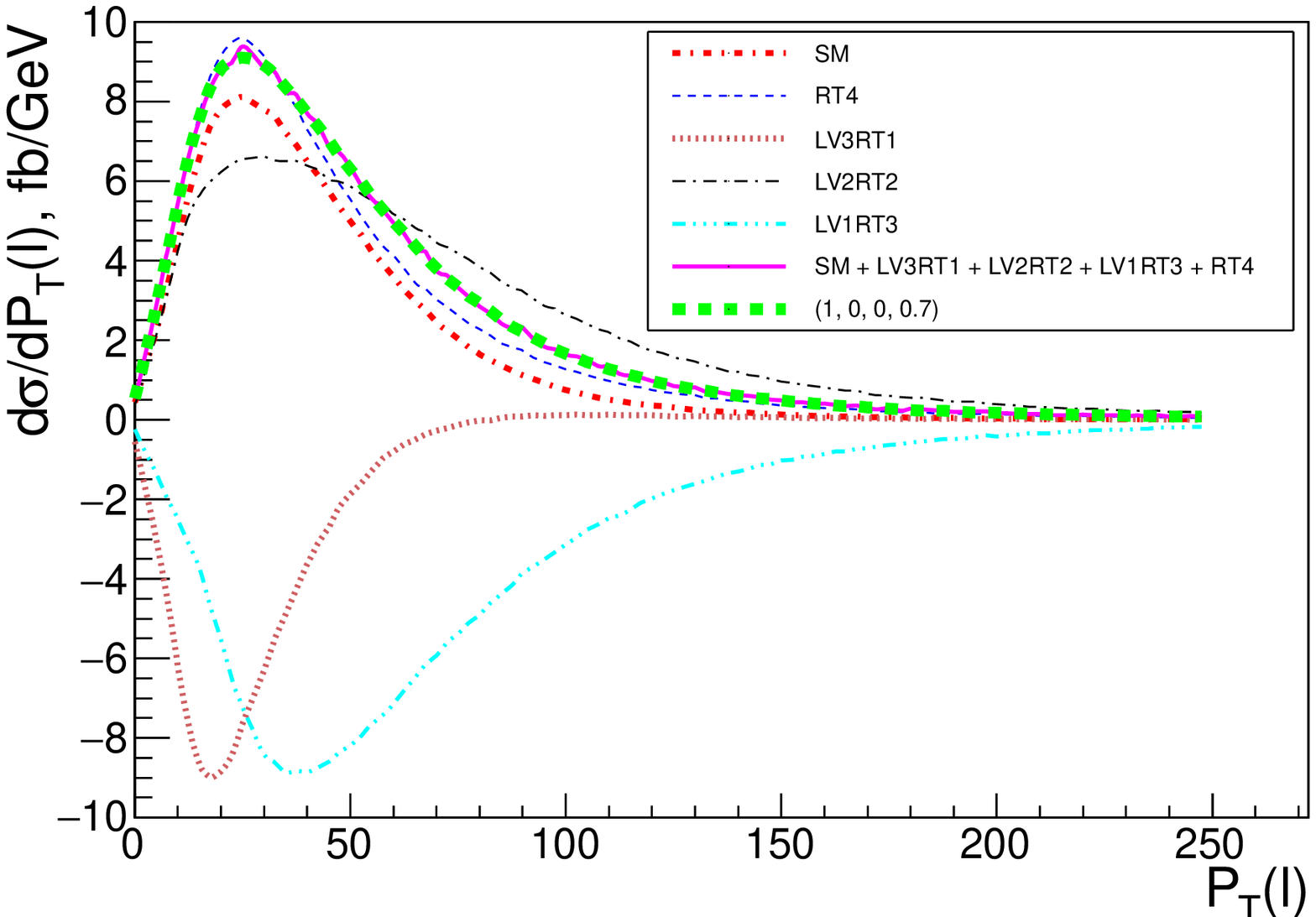}
\includegraphics[width=.45\textwidth]{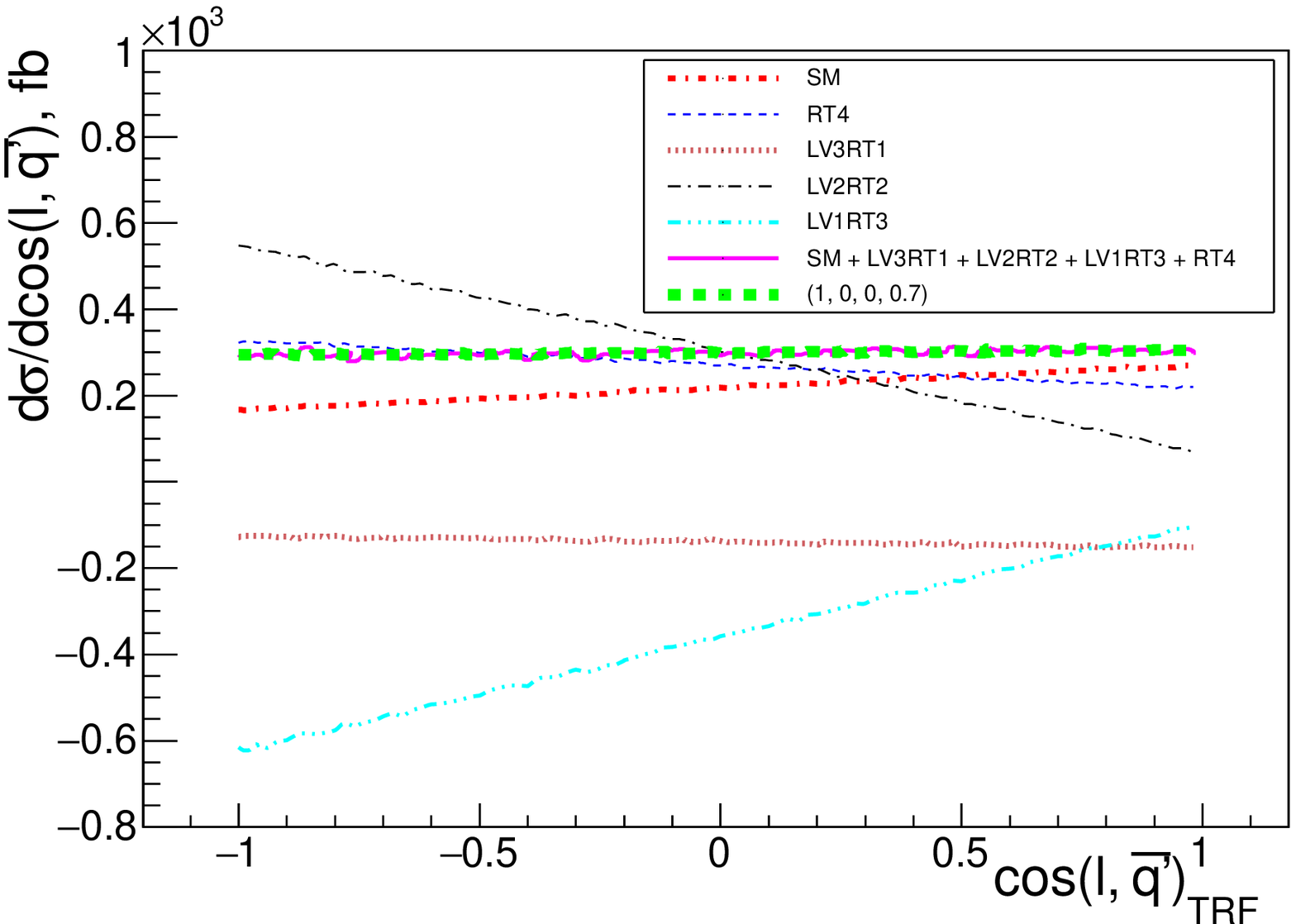}
\caption {   The distributions for the transverse momentum of lepton from top quark decay in center of mass rest frame (left plot) and the cosine of angle between lepton from the top quark decay and down-type quark in the top quark rest frame (right plot) for the process $pp \rightarrow t (\bar {\nu_{l}},{l}, \bar {b}), q, \bar {q'} $ (tW-channel single top quark production) for $(f_{LV}, f_{RT})$ scenario.  \label{LVRT_tW_ch_plots} }
\end{figure*}

For the illustration of the method the case with $f_{LV}=1.0$ 
and $f_{RT}=0.8$, $f_{RT}=0.7$ values of the couplings in \wtb~vertex is simulated 
with the combination of the described minimal set of event samples. The results are shown in Figs.~\ref{LVRT_s_ch_plots}-\ref{LVRT_tW_ch_plots}. For example, in Fig.\ref{LVRT_s_ch_plots} (left plot) the curve ``1, 0, 0, 0.8`` shows the distribution of the transverse momenta of the lepton from the top quark decay for the case with $f_{LV}=1.0$ and $f_{RT}=0.8$ values of the anomalous couplings in \wtb~vertex. The curve "SM" shows  the same  distribution of  $f_{LV}^4\cdot(\rm LV4)$ set of events and 
the curve ``RT4'' shows  the same  distribution of  $f_{RT}^4\cdot(\rm RT4)$ set of events.
The curves ``LV3RT1'', ``LV2RT2'', and ``LV1RT3'' show the distributions of the event sets which correspond to the parts of~(\ref{me_sq_lvrt_pr_decay}) and the squared diagrams with SM W boson and $W_{\rm subs}$,  multiplied by the factors $f_{LV}^3 f_{RT}$, $f_{LV}^2 f_{RT}^2$, and $f_{LV} f_{RT}^3$. 
The agreement of  curve ``1, 0, 0, 0.8`` and the sum of the five event sets are shown in Figs.~\ref{LVRT_s_ch_plots}-\ref{LVRT_tW_ch_plots}.

Since the cross terms between
 $f_{\rm T}^{\rm L}$ and $f_{\rm T}^{\rm R}$ or  $f_{\rm V}^{\rm R}$ 
and $f_{\rm T}^{\rm R}$ couplings are suppressed, it is possible 
to use the event samples described above to simulate kinematics with 
three-dimensional variation of the 
$f_{\rm V}^{\rm L}$, $f_{\rm T}^{\rm L}$, $f_{\rm T}^{\rm R}$
 or $f_{\rm V}^{\rm L}$, $f_{\rm V}^{\rm R}$, $f_{\rm T}^{\rm R}$ couplings.
 For the simulation of kinematics with all four couplings additional samples 
are needed.

One should note that the contributions with odd powers of anomalous couplings are negative for the positive values of the couplings as shown in Figs.~\ref{LVRT_s_ch_plots}-\ref{LVRT_tW_ch_plots} for ``LV3RT1'' and ``LV1RT3'' curves, and 
this fact has to be taken into account while generating events for an experimental analysis.

\subsection{Application of the method to the experiment}
\label{subsec:experiment}
The application of this method to the experimental search for anomalous Wtb couplings in the single top quark production processes is straightforward. Usually, two assumptions are accepted, neglecting the b quark mass comparing to the top quark mass and the narrow width approximation for the top quark.
The first assumption leads to  the simple expression of the cross section dependence on the anomalous couplings (Eq.~\ref{xsection}); the non-zero b-quark mass leads to the presence of the cross terms in Eq.~\ref{xsection} not only for  $(f_{LV}, f_{RT})$ and $(f_{RV}, f_{LT})$ couplings but for all pairs of the couplings. However these additional terms are suppressed by the factor of $(\frac{m_b}{m_{top}})^2$ and are neglected for the experimental tasks. The using of the narrow width approximation is also reasonable because even for the anomalous couplings much larger than the current limits the top quark width is much smaller than the top quark mass. 

It was shown in the previous sections that the minimal number of event samples is different in different scenarious. Namely, for the simple scenarios  $(f_{LV}, f_{RV})$ or  $(f_{LV}, f_{LT})$ three 
event samples (``LV4'', ``LV2RV2'', ``RV4'' or ``LV4'', ``LV2LT2'', ``LT4'') are necessary while for the third scenario $(f_{LV}, f_{RT})$  five event samples (``LV4'', ``LV3RT1'', ``LV2RT2'', ``LV1RT3'', ``RT4'') have to be generated.
   In practice, each event sample is generated using its own top quark width according to the coupling values for the sample. For example in the scenario $(f_{LV}, f_{LT})$ in samples ``LV4'', ``LV2LT2'', ``LT4'' the top quark total widths  $w_{\rm tot}(1,0,0,0)$, $w_{\rm tot}(1,0,1,0)$, $w_{\rm tot}(0,0,1,0)$ are used respectively.
However in order to get the self consistent sum of the distributions followed from the event samples for the scenario $(f_{LV}, f_{LT})$ one should multiply each distribution following from the particular event sample by the reweighting factors as given in Eq.~(\ref{sigma_lvlt_modelling_corrections}):
\begin{equation}
\begin{split}
(f_{LV},~0,~f_{LT},~0)  =& \left( f_{LV} \right)^{4}  \frac{w_{\rm tot}(1,0,0,0)}{w_{\rm tot}(f_{LV},0,f_{LT},0)} \cdot {(\rm LV4)}  \\ & + \left( f_{LV} \right)^{2}\left( f_{LT} \right)^{2}  \frac{w_{\rm tot}(1,0,1,0)}{w_{\rm tot}(f_{LV},0,f_{LT},0)} \cdot {(\rm LV2LT2)} \\ & + \left( f_{LT} \right)^{4} \frac{w_{\rm tot}(0,0,1,0)}{w_{\rm tot}(f_{LV},0,f_{LT},0)} \cdot  {(\rm LT4)}.
\label{sigma_lvlt_modelling_corrections}
\end{split}
\end{equation}

\begin{figure*}[h]
\centering
\includegraphics[width=.45\textwidth]{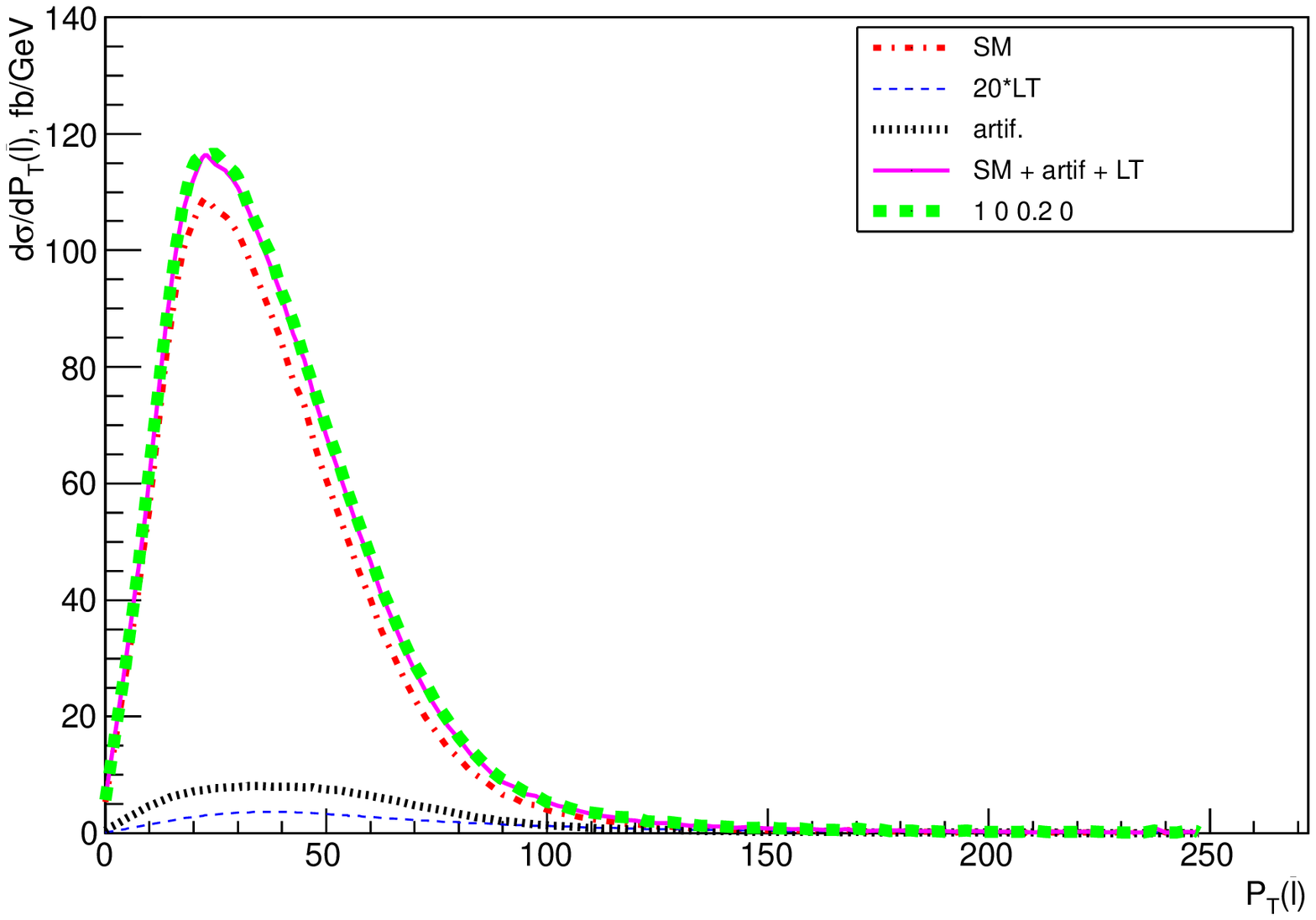}
\includegraphics[width=.45\textwidth]{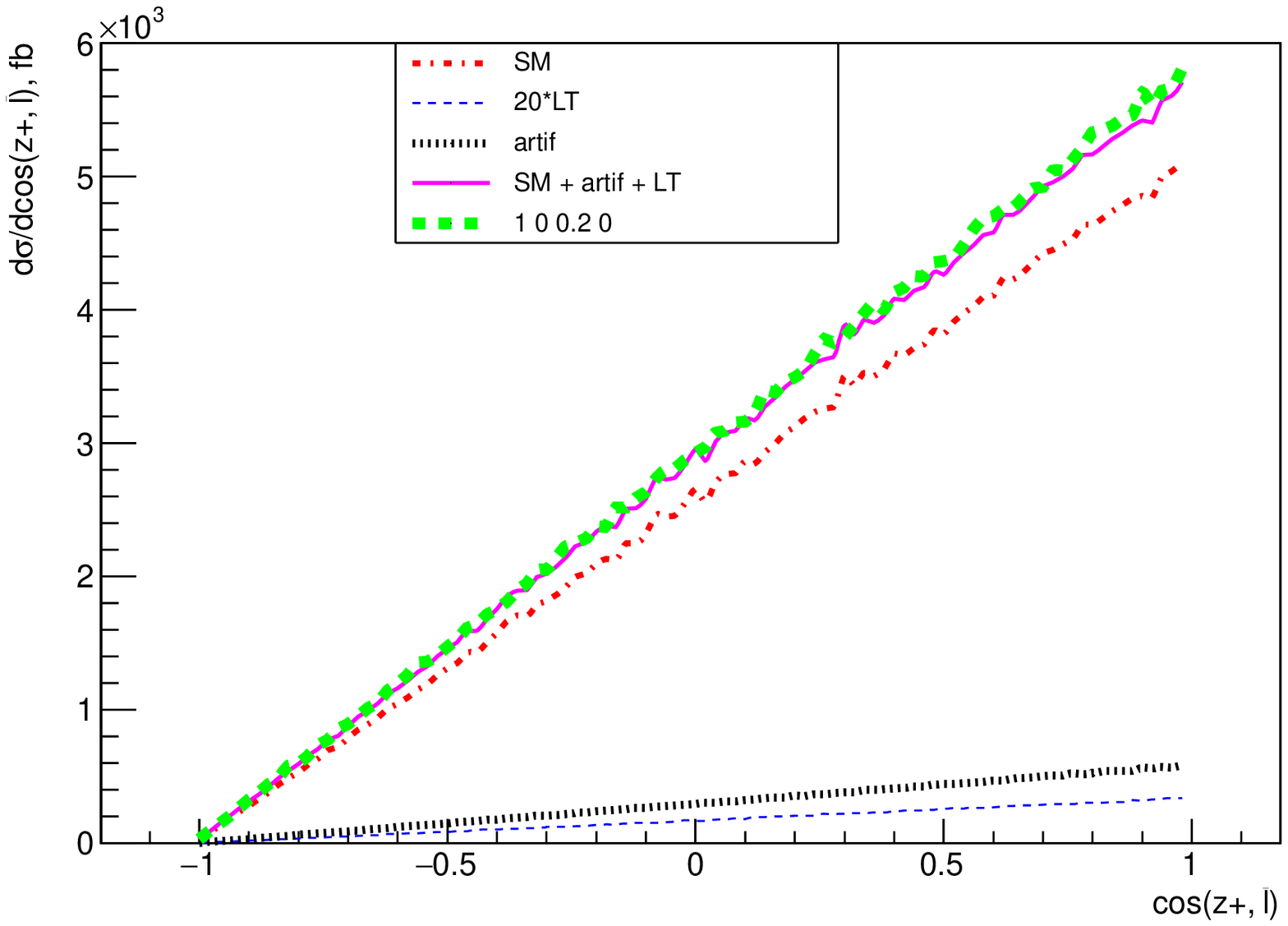}
\caption { The distributions for the transverse momentum of top quark in center of mass rest frame (left plot) and the cosine of angle between lepton from the W boson decay from top quark and light quark in the top quark rest frame (right plot) for the process $pp \rightarrow t (\nu_{l},\bar {l}, b) q$ (t-channel single top quark production) for $(f_{LV}, f_{LT})$ scenario without the assumption of massless b-quark and with the adjusted factors which are related to different values of top quark widths with anomalous couplings are present.   \label{LVLT_corr_t_ch_plots} }
\end{figure*}

Figure~\ref{LVLT_corr_t_ch_plots} demonstrates that the factors in Eq.~(\ref{sigma_lvlt_modelling_corrections}) should be included to reproduce correctly the total result using the sum of the individual contributions. The formula~(\ref{sigma_lvlt_modelling_corrections}) shows how to get the distribution from the event set with arbitrary values of the anomalous couplings from the distributions following from the basic event sets with anomalous couplings taken to be one or zero. In the example for the Fig.~\ref{LVLT_corr_t_ch_plots} non-zero b-quark mass was used and, as expected, the influence of non-zero b-quark mass is negligible.

For the simulation of the event samples with odd powers of $f_{RT}$ coupling the negative value of the coupling can be chosen to have positive cross section of the event sample. Then all possible values of the anomalous couplings can be considered in the statistical analysis according to Eq.~(\ref{me_sq_lvrt_pr_decay}) for the full matrix element. 

In the most general scenario when all anomalous couplings are taken to be non-zero and neglecting the b quark mass, the minimal number of event samples is equal to twelve:  ``LV4'', ``LV2RV2'', ``RV4'', ``LV2LT2'', ``LT4'', ``LV3RT1'', ``LV2RT2'', ``LV1RT3'', ``RT4'', ``RV3LT1'', ``RV2LT2''~~ and ~~``RV1LT3''\footnote{The phenomenological analysis including correlations between anomalous couplings is given in \cite{Cao:2015doa}}. Such event samples with unit values for the anomalous couplings allows to use same samples in different scenarios,  and they form a minimal basis set of event samples. 
The common event sample is constructed from the basis set of samples in the same way as given in Eq.~(\ref{sigma_lvlt_modelling_corrections}) with the total top quark width $w_{\rm tot}(f_{LV},f_{RV},f_{LT},f_{RT})$ in the denominator of the reweighting factors.

The event sets of the single top quark production including anomalous Wtb couplings were prepared as described above for the LHC energies and uploaded into the open access Monte-Carlo simulated event database~\cite{Belov:2007qg}.

One should also notice, since the $W_{\rm subs}$ interacts with other particles by electroweak 
forces, the introduction of this particle is not affected by the NLO QCD 
corrections, and therefore the method is applicable at the NLO QCD level as well.

\section{Conclusion}
\label{sec:Conclusion}

In the paper a new method of modeling the events with anomalous fermion-boson couplings is presented. The method is based on introducing subsidiary vector fields in addition to the SM gauge field in the unitary gauge. The subsidiary fields have the same masses and all the couplings to fermions as the SM gauge field except the couplings to the fermion with anomalous interaction. The coupling of the subsidiary field to that fermion is the anomalous coupling. In case of several anomalous couplings contributing simultaneously to the production and to the decay  as well as to various interference terms the method performs correct simulation of the dependence of kinematical  distributions on anomalous couplings with the minimal set of event samples. The method allows to perform simulations in two different approaches keeping only the linear order contributions or keeping higher order contributions in anomalous couplings. The first approach is motivated by the effective field theory (EFT) in which the only leading  $1/{\Lambda}^2$ contributions are taken into account. In the second approach higher orders in $1/{\Lambda}^2$ are also taken into account as appeared in direct matrix element computations. Since each of the anomalous coupling is associated with corresponding subsidiary field it is very easy to keep only needed contribution by removing all not needed diagrams from the amplitude or squared diagrams from the matrix element squared. In fact, it is very instructive to use both approaches simultaneously since a comparison of results in two cases allows to understand a region of applicability (EFT) in the anomalous parameter space. 

The method allows to simplify significantly realistic analyses by generating  
only the minimum number of the event samples with the unity values of the anomalous couplings. The method is very easy to implement in different computing codes, as was done in this study using CompHEP. The proposed method works for arbitrary widths of the fermion resonances. However the simple formula to rescale contributions from different sets of events (such as Eq.~(\ref{sigma_lvlt_modelling_corrections}) works only in the narrow width approximation.

Practical use of proposed method is demonstrated in an example of the single top quark production  processes with anomalous \wtb~couplings.  In our demonstration we focused on more difficult for the analysis approach computing Feynman diagrams with non-linear behaviour of the anomalous couplings. In this case the terms with higher dimensions on $1/{\Lambda}^2$ arise due to the multiplication of the production and decay parts of the processes both depending on anomalous couplings and the presence of the total top quark width in the denominator. One should stress that if one considers only the leading terms of the order of $1/{\Lambda}^2$ one needs not only to keep leading terms in numerator of diagrams but also to expand the total top quark width in the denominator and to take into account the terms with the dimension of $1/{\Lambda}^2$ in the overall expansion. 

\section{Acknowledgements}
\label{sec:acknownlegements}

The authors are grateful to R.~Schwienhorst, H.~Prosper and M.~Dubinin as well as many colleagues from \dzero and \cms single top groups for useful discussions and comments. The work was supported by grant 16-12-10280 of the Russian Science Foundation.

\appendix
\section{Cross sections for the single top quark production processes with the anomalous \wtb~couplings} \label{Appendix}
The cross sections of the single top quark production processes for s- and t-channels as well as the cross section of the  top quark production in association with a W boson in the presence of all anomalous \wtb~couplings from (\ref{anom_wtb_eq_lagrangian}) have the following expressions:

for s-channel:
\begin{align}
\sigma(\hat{s})_{u\bar{d}\to t\bar{b}} = ~&\frac{\pi\cdot V_{ud}^2\cdot \alpha^2}{24\sin^4{\Theta_W}}\cdot\frac{\beta^4\cdot\hat{s}}{(\hat{s} - m_W^2)^2}~\times \\ \nonumber &
\big[ ~~(3 - \beta^2)\cdot\big(f_{LV}^2 + f_{RV}^2\big) \\ \nonumber &
+ (3 - 2\beta^2)\cdot \frac{\hat{s}}{m_W^2}\cdot\big(f_{LT}^2 + f_{RT}^2\big) \\ \nonumber &
- \frac{6m_t}{m_W}\cdot\big(f_{LV}\cdot f_{RT} + f_{RV}\cdot f_{LT}\big)~~\big]~~~~~~~~~~~~~~~~~~ \nonumber &
\end{align}
where:
\begin{align}
\nonumber &\beta^2 = 1-\frac{m_t^2}{\hat{s}};~~~~~~~~~~~~~~~~~~~~~~~~~~~~~~~~~~~~~~~~~~~~~~~~~~~~~~~~~~~~~~~~
\end{align}

for t-channel:
\begin{align}
&\sigma(\hat{s})_{ub\to td} = \frac{\pi\cdot V_{ud}^2\cdot\alpha^2}{4\cdot \hat{s}\cdot\sin^4{\Theta_W}}~\times\\ \nonumber &
\big[ ~~c_0 c_p \beta^4\cdot f_{LV}^2 \\ \nonumber &
+ \left( - (1 + c_1)\cdot ln(a_1)~ + ~(2 + c_0)\cdot\beta^2\right)\cdot f_{RV}^2 \\ \nonumber &
+ \big(~~(2 + c_0)\cdot ln(a_1)~ - ~(1 + c_1)\cdot c_0c_p\beta^2\big)\cdot f_{RT}^2 \\ \nonumber &
+ \left( c_1\cdot ln(a_1)~ - ~2\beta^2\right)\cdot c_0\beta^2\cdot f_{LT}^2 \\ \nonumber &
+ \frac{2m_t}{m_W}\cdot \left( (-ln(a_1) + c_p\beta^2)\cdot f_{LV}\cdot f_{RT}\right)  \\ \nonumber &
+ \frac{2m_t}{m_W}\cdot \left(( c_1\cdot ln(a_1) - 2\beta^2\big)\cdot f_{RV}\cdot f_{LT}\right)~~\big]  \nonumber &
\end{align}
where:
\begin{align}
\nonumber &\beta^2 = 1-\frac{m_t^2}{\hat{s}},~~~a_1 = 1 + \frac{\beta^2\hat{s}}{m_W^2},~~~c_p = \frac{\hat{s}}{(\hat{s} - m_t^2 +  m_W^2)},\\ \nonumber &~~~c_0 = \frac{\hat{s}}{m_W^2},~~~ 
c_1 = \frac{2m_W^2}{\hat{s}} + \beta^2;
\end{align}

for tW-channel:
\begin{align}
&\sigma(\hat{s})_{bg\to tW^-}= \frac{3\cdot\pi\cdot\alpha\cdot\alpha_s}{32\cdot m_W^2\cdot\sin^2{\Theta_W}}~\times\\ \nonumber 
 & \big[\big( (\frac{c_5}{2} + c_2 c_5 + c_2^2 c_5)\cdot ln(a_2)~ \\ \nonumber & + ~ \big(- c_2-c_3 + c_2 c_5\big)\cdot \frac{\delta\beta}{4} \big)\cdot\left(f_{LV}^2 + f_{RV}^2\right) \\ \nonumber 
& + \big( (-\frac{c_3}{2} + c_2 c_4 + c_2^2 c_4 )\cdot ln(a_2)~ \\ \nonumber & + ~ (4 - c_3 + c_2 c_4)\cdot \frac{\delta\beta}{4} \big)\cdot \left(f_{LT}^2 + f_{RT}^2\right) \\ \nonumber 
& + \big((\frac{1}{6} - c_2 - c_2^2)\cdot ln(a_2)~ \\ \nonumber & + ~ (\frac{1}{3} - c_2)\cdot \frac{\delta\beta}{4} \big)\cdot \frac{6 m_t\cdot m_W}{\hat{s}}\cdot \big(f_{LV}\cdot f_{RT} + f_{RV}\cdot f_{LT}\big)\big] \nonumber 
\end{align}
where:
\begin{align}
\nonumber &c_2 = \frac{m_t^2 - m_W^2}{\hat{s}}, ~c_3 = \frac{2m_t^2 - m_W^2}{\hat{s}},~c_4 = \frac{2m_t^2 + m_W^2}{\hat{s}},~~~\\ \nonumber & c_5 = \frac{m_t^2 + 2m_W^2}{\hat{s}}, a_2= \frac{\hat{s} + m_t^2 - m_W^2 + \hat{s}\cdot\delta\beta}{\hat{s} + m_t^2 - m_W^2 - \hat{s}\cdot\delta\beta},~~~\\ \nonumber & \delta=\sqrt{1 - \frac{(m_t - m_W)^2}{\hat{s}}},~~~ \beta=\sqrt{1 - \frac{(m_t + m_W)^2}{\hat{s}}}. 
\end{align}


\end{document}